%% file: IronSpreadGC2.tex
\documentclass[fleqn,usenatbib]{mnras}

\usepackage[T1]{fontenc}

\DeclareRobustCommand{\VAN}[3]{#2}
\let\VANthebibliography\thebibliography
\def\thebibliography{\DeclareRobustCommand{\VAN}[3]{##3}\VANthebibliography}

\usepackage{graphicx}	
\usepackage{placeins}    
\usepackage{siunitx}		
\usepackage{amsmath}	
\usepackage{amssymb}	
\usepackage{aas_macros}
\usepackage{enumitem} 
\usepackage[shortcuts,acronyms]{glossaries-extra}		

\newabbr{GC}{GC}{globular cluster}
\newabbr[longplural = core collapse supernovae, shortplural = CCSNe]{SN}{CCSN}{core collapse supernova}
\newabbr{IMF}{IMF}{initial mass function}
\newabbr{SF}{SF}{star formation}
\newabbr{SFE}{SFE}{star formation efficiency}
\newabbr{SFR}{SFR}{star formation rate}
\newabbr{MW}{MW}{Milky Way}
\newabbr{AGB}{AGB}{asymptotic giant branch}
\newabbr{SMBH}{SMBH}{super-massive black hole}
\newabbr{BH}{BH}{black hole}
\newabbr{NS}{NS}{neutron star}

\title[Top-heavy IMFs in GCs]{The giants that were born swiftly - Implications of the top-heavy stellar initial mass function on the birth conditions of globular clusters}

\author[H. Wirth et al.]{
Henriette Wirth$^{1}$\thanks{E-mail: wirth@sirrah.troja.mff.cuni.cz (HW)},
Pavel Kroupa$^{1,2}$,
Jaroslav Haas$^1$,
Tereza Jerabkova$^{3}$,
\vspace{5pt}\\
\rm \Large{Zhiqiang Yan$^{1,4}$
and Ladislav Šubr$^1$}
\\
$^{1}$Charles University, Faculty of Mathematics and Physics, Astronomical Institute, V Hole\v{s}ovi\v{c}kách 2, Praha, CZ-18000, Czech Republic\\
$^{2}$Helmholtz Institut für Strahlen und Kernphysik, Universität Bonn, Nussallee 1416, 53115 Bonn, Germany\\
$^{3}$European Southern Observatory, Karl-Schwarzschild-Strasse 2, 85748 Garching bei M\"unchen\\
$^{4}$School of Astronomy and Space Science, Nanjing University, Nanjing, 210023, China
}

\date{Accepted XXX. Received YYY; in original form ZZZ}

\pubyear{2021}

\begin{document}
\label{firstpage}
\pagerange{\pageref{firstpage}--\pageref{lastpage}}
\maketitle

\begin{abstract}
Recent results suggest that the \ac{IMF} of \acp{GC} is metallicity and density dependent.
Here it is studied how this variation affects the initial masses and the numbers of \acp{SN} required to reproduce the observed iron spreads in \acp{GC}.
The \acp{IMF} of all of the investigated \acp{GC} were top-heavy implying larger initial masses compared to previous results computed assuming an invariant canonical \ac{IMF}.
This leads to more \acp{SN} being required to explain the observed iron abundance spreads.
The results imply that the more massive \acp{GC} formed at smaller Galactocentric radii, possibly suggesting in-situ formation of the population II halo.
The time until \ac{SF} ended within a proto-\ac{GC} is computed to be 3.5 - 4 Myr, being slightly shorter than the 4 Myr obtained using the canonical \ac{IMF}.
Therefore, the impact of the \ac{IMF} on the time for which \ac{SF} lasts is small.
\end{abstract}


\glsresetall



\section{Introduction}

While most \acp{GC} have been observed to be homogeneous with respect to iron \citep{2009A&A...508..695C, 2015ApJ...809..128M, 2021arXiv210307014M}, there is evidence for iron abundance spreads in some of them \citep{2000ApJ...534L..83P,2009Natur.462..483F,2016MNRAS.457...51L,2018ApJ...859...81M, 2019ApJS..245....5B,2022arXiv220503323L}.
This spread in iron is believed to be caused by pollution through \acp{SN}, however, it is unclear at which stage of \ac{GC} development these \acp{SN} exploded \citep{2016MNRAS.458.2122D,2021MNRAS.506.4131W,2021MNRAS.506.5951L}.
In \citet[hereinafter Paper I]{2021MNRAS.506.4131W} a novel method to estimate the number of \acp{SN} required to explain the iron abundance spread in a \ac{GC} assuming that the iron spread is caused by \ac{SN} was introduced.
This approach was based on a model in which \acp{GC} without an iron spread arise if \ac{SF} stops before the first \ac{SN} can pollute the gas in the \ac{GC}.
If \ac{SF} continues past this point, the \acp{SN} will pollute the surrounding gas with iron, such that more iron rich stars form.
Paper I shows how the number of \acp{SN} contributing to this iron rich population and the time after which \ac{SF} ends can be computed using the iron spread catalogue from \cite{2019ApJS..245....5B}.

An integral part of the calculations in Paper I was to compute the initial masses of the \acp{GC}.
To this end we used a set of equations which \cite{2003MNRAS.340..227B} derived from N-body simulations.
The initial mass of a \ac{GC} is, like other \ac{GC} parameters, related to the assumed \ac{IMF} \citep{2006A&A...458..135P,2012MNRAS.422.2246M,2017MNRAS.471.2242B,2017A&A...608A..53J,2018ApJ...857..132K,2018Sci...359...69S,2021arXiv210304997C}.
In Paper I, the canonical \ac{IMF} \citep{2001MNRAS.322..231K,2013pss5.book..115K} was assumed for all \acp{GC}.
However, several theoretical \citep{2022MNRAS.509.1959S} and observational \citep{2012MNRAS.422.2246M,2018Natur.558..260Z,2021A&A...655A..19Y,2022arXiv220303276P} studies have found that the \ac{IMF} does change depending on the initial density and metallicity of a stellar population.

To determine the \ac{IMF} of a \ac{GC} is challenging since the mass function of a \ac{GC} changes over time.
This is mainly due to two effects: the fact that massive stars die first \citep[see e.g.][]{1998A&A...334..505P,2001A&A...370..194M} and the loss of stars due to energy-equipartion driven evaporation and ejection \citep{2001ApJ...561..751F,2003gmbp.book.....H,2015MNRAS.453.3278W}.
Additionally, mergers can play a role \citep{2022MNRAS.512.2936K}.

Despite these obstacles, several attempts have been made to compute the \ac{IMF} of a \ac{GC}.
Only recently, \cite{2021arXiv210802217W} suggested to use the mass function observed at the ends of the tidal tails of a \ac{GC} to determine the \ac{IMF}.
They do mention, however, that even \textit{Gaia} will be unable to detect low-mass stars ($< 0.7 M_\odot$ for an isochrone and distance of GD-1 and Pal 5) and this does not solve the problem that stars more massive then $\approx 0.8 M_\odot$ already evolved off the main sequence \citep{1998A&A...334..505P,2001A&A...370..194M}, making them undetectable.
As \citet{2021A&A...647A.137J} pointed out, the detection of tidal tails is challenging in general.
They suggest a new method, the compact convergent point method, to extract the tidal tails.

\cite{2001MNRAS.322..231K} combined constrains from different studies into what is known as the canonical \ac{IMF} and concluded that the existing data may be indicating a systematic variation with metallicity.
\cite{2002Sci...295...82K} then formulated evidence for the \ac{IMF} becoming more bottom-light (fewer low mass stars) with decreasing iron abundance.
Additionally, \cite{2012MNRAS.422.2246M} found that the slope of the upper end of the \ac{IMF} (for stars with a mass $m > 1 M_\odot$) depends on the initial gas cloud density and the iron abundance.
They compared a set of numerical simulations to observations to determine the connection between the initial mass function and the properties of the gas clouds the \acp{GC} formed out of.
\cite{2021A&A...655A..19Y} later found a systematic variation in the low-mass part of the \ac{IMF} as well.

The purpose of this paper is, thus, to investigate how a varying \ac{IMF} affects the initial mass computed for the \acp{GC} and the number of \acp{SN} required to reach the observed spread of $[Fe/H]$.
The results of this study will be compared to those in Paper I.
Sec. \ref{sec_methods} explains how the systematic variation of the \ac{IMF}, as documented in \cite{2021A&A...655A..19Y} is applied to the computation of the initial masses of the \acp{GC}.
Additionally the method to compute the time after which \ac{SF} ends is explained.
Secs. \ref{sec_res} and \ref{sec_disc} contain the results and discussion, respectively, and finally we will draw our conclusions in Sec. \ref{sec_concl}.

\section{Methods}
\label{sec_methods}

\subsection{Cluster initial mass as a function of the IMF}
\label{sec_metIMF_ClM}

In Paper I the results from \cite{2003MNRAS.340..227B} were used to compute the initial stellar mass of the cluster, $M_{\rm ini}$.
Note that this is the mass after revirialisation following a \ac{GC}'s gas expulsion.
\cite{2003MNRAS.340..227B} use the connection between dissolution time, $t_{\rm diss}$, half-mass relaxation time, $t_{\rm rh}$, and crossing time, $t_{\rm cross}$, derived in \cite{2001MNRAS.325.1323B}: $t_{\rm diss} = k t_{\rm rh}^{x} t_{\rm cross}^{1-x}$.
They combine this with the results of their simulations to formulate an equation for the dissolution time of clusters and another one for the evolution of their mass.
In Paper I, these findings were combined into one implicit equations that can be solved for $M_{\rm ini}$ numerically:
\begin{align}
    \label{eq_Mini}
    0 &= \beta \left[ \frac{\frac{M_{\rm ini}}{\left<m\right>}}{\ln( \gamma \frac{M_{\rm ini}}{\left<m\right>} )} \right]^x \frac{R_{\rm ap}}{\mathrm{kpc}} (1 - e) \frac{1-\frac{M(t)}{p_{\rm SF} M_{\rm ini}}}{\frac{t}{\mathrm{Myr}}} - 1.
\end{align}
Note that for consistency the terms are written down in the same order as in Paper I.
$\beta$ and $x$ are fitting parameters which depend on the King concentration parameter, $W_0$.
$\gamma = 0.02$ is the Coulomb logarithm.
$R_{\rm ap}$ and $e$ are the apocentre and eccentricity of the \ac{GC}'s orbit, respectively.
$M(t)$ is the current mass of the \ac{GC} and $t$ its age.
Analogous to Paper I, the age is assumed to be 12 Gyr in accordance with literature values \citep{2010ApJ...708..698D,2019MNRAS.490..491U,2021arXiv210908708C}.
The mean stellar mass $\left< m \right>$ and the portion of $M_{\rm ini}$ lost through stellar evolution, $p_{\rm SF}$, will here be computed depending on the \ac{IMF}.
This is assumed to happen instantaneously \citep{2003MNRAS.340..227B}.
Paper I follows \cite{2003MNRAS.340..227B} and assumes $p_{\rm SF} = 0.7$, which means that 30\% of the initial mass was assumed to be lost due to stellar evolution for an invariant canonical \ac{IMF}.

\subsubsection{Dependence of the IMF on initial cluster mass}
\label{sec_metClM_IMF}

\begin{figure}
    \includegraphics{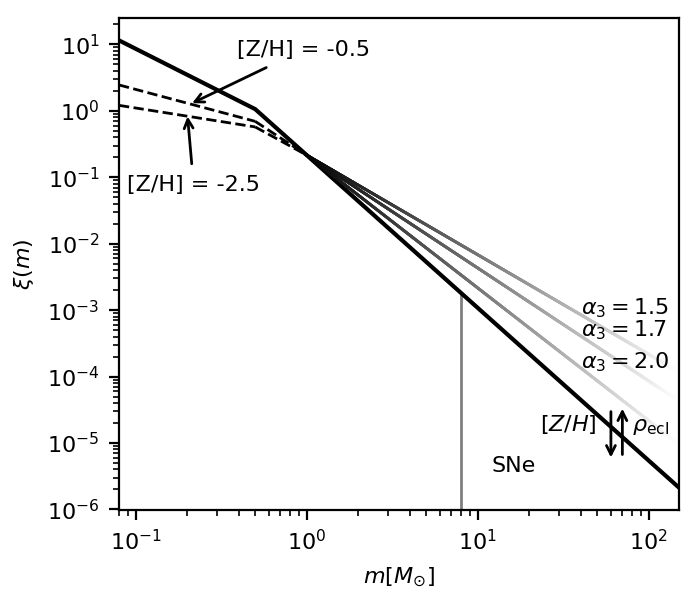}
    \caption{The canonical mass function $\xi(x)$ normalized to one (the area below the curve is the number fraction of stars).
    The mass beyond which stars are expected to explode in a CCSN ($> 8 M_\odot$) is marked with a vertical grey line.
    Additionally, changes to the low-mass part of the mass function depending on $[Z/H]$ (dashed line, Eqs. \ref{eq_alpha1} and \ref{eq_alpha2}), and the metallicity- and density-dependent $\alpha_3$ (faded line, Eqs. \ref{eq_alpha3} to \ref{eq_y}) are shown.}
    \label{fig_massDens}
\end{figure}

The \ac{IMF}, $\xi (m)$, is defined through the number of stars born together in the mass interval $[m,m + dm]$, $dN = \xi(m) dm$.
\cite{2012MNRAS.422.2246M} derived the correlation between the \ac{IMF}, initial central density and metallicity of a \ac{GC} by comparing numerical results based on gas-expulsion simulation with observations.
Using more recent data, \cite{2021A&A...655A..19Y} rewrote the variations in all parts of the \ac{IMF} as a function of $[Z/H]$ rather than $[Fe/H]$.
They formulated the \ac{IMF}, $\xi(m) = k_i m^{-\alpha_i}$, as:
\begin{align}
    \label{eq_alpha1}
    \alpha_1 &= 1.3 + \Delta\alpha (10^{[Z/H]_{\rm ini}}-1) Z_\odot,  &0.08 \leq &\frac{m}{M_\odot} \leq 0.5,\\
    \label{eq_alpha2}
    \alpha_2 &= 2.3 + \Delta\alpha (10^{[Z/H]_{\rm ini}}-1) Z_\odot,  &0.5 < &\frac{m}{M_\odot} \leq 1,\\
    \label{eq_alpha3}
    \alpha_3 &= \left\lbrace \begin{aligned} &2.3, & y &< -0.87, \\
    &-0.41 y + 1.94, & y &\geq -0.87, \end{aligned} \right. & 1 < &\frac{m}{M_\odot},\\
    \label{eq_y}
    y &= \rlap{$-0.14 [Z/H]_{\rm ini} + 0.99 \log_{10} \left( \rho_{\rm gas} / (10^6 M_\odot {\rm pc^{-3}}) \right),$}
\end{align}
with the initial metallicity of the \ac{GC}, $[Z/H]_{\rm ini}$, and its cloud core density, $\rho_{\rm gas}$.
$Z_\odot = 0.0142$ is the metal mass fraction in the Sun, $\Delta\alpha = 63$ is a constant and $k_i$ is a constant assuring the function is continuous and normalized.

Following \cite{2011MNRAS.413.2943F}, $[Fe/H]$ is converted to the total metallicity, $[Z/H]$, using $[Z/H] = [Fe/H] + [\alpha /H]$.
The value of $[\alpha /H] = 0.3$ is typical for Galactic \acp{GC} \citep{1996PASP..108..900C,2011MNRAS.413.2943F}.
The dependency of the \ac{IMF} on the metalicity and $\rho_{\rm gas}$ is not arbitrary, but has been extracted from data such that the \ac{IMF} is consistent with:
\begin{itemize}[wide, labelwidth=!, labelindent=\parindent]
    \item direct observational constraints from the Galactic field population and a large range of star clusters \citep{2002Sci...295...82K,2012MNRAS.422.2246M},
    \item the observed dynamical mass-to-light ratios and numbers of low-mass X-ray binary systems in ultra-compact dwarf galaxies \citep{2009MNRAS.394.1529D,2012ApJ...747...72D},
    \item the synthesised galaxy-wide \acp{IMF} of galaxies of different types \citep{2017A&A...607A.126Y}, as well as
    \item galaxy chemical evolution studies \citep{2021A&A...655A..19Y}.
\end{itemize}
All the aforementioned IMF constraints have uncertainties, but the self-consistency-check attained by calculating galaxy-wide stellar populations \citep{1993MNRAS.262..545K,2017A&A...607A.126Y,2018A&A...620A..39J,2020A&A...637A..68Y,2021A&A...655A..19Y} suggests that any true variation is comparable to the above formulation.

\cite{2012A&A...543A...8M} found the following correlation between the density, $\rho_{\rm ecl}$, and the mass, $M_{\rm ecl}$, of the stars in the still gas embedded clusters by fitting the densities and masses of several datasets (see their fig. 6):
\begin{align}
    \label{eq_rho}
    \log_{10} \frac{\rho_{\rm ecl}}{M_\odot {\rm pc^{-3}}} = a \log_{10} \frac{M_{\rm ecl}}{M_\odot} + b,
\end{align}
with $a = 0.61 \pm 0.13$ and $b = 2.08 \pm 0.69$.
The gas density of the cloud is then $\rho_{\rm gas} = \frac{\rho_{\rm ecl}}{\epsilon}$, where $\epsilon$ is the \ac{SFE}.
In agreement with Paper I, $\epsilon = 0.3$ is assumed \citep[see e.g.][]{2003ARA&A..41...57L,2014prpl.conf...27A,2016AJ....151....5M,2018ASSL..424..143B}.
Based on the gas expulsion computations by \cite{2017A&A...600A..49B} it is assumed that a \ac{GC} looses a negligible fraction of its stars when its residual gas is expelled, i.e. $M_{\rm ini} = M_{\rm ecl}$ to a good approximation.
With these calculations, the IMF can now be expressed in dependence of to $M_{\rm ini}$ and $[Fe/H]$.

Possible forms of the \ac{IMF} are illustrated in Fig. \ref{fig_massDens}.
The canonical \ac{IMF} is shown by the thick black line.
The changes in $\alpha_1$ and $\alpha_2$ depending on $[Fe/H]$ are made visible by dashed lines.
Additionally, the shape of the \ac{IMF} for different $\alpha_3$ are shown.
The \acp{IMF} becomes more bottom-light (fewer low mass stars per star formed) with decreasing $[Fe/H]$.
Simultaneously, a lower $[Fe/H]$ and a higher gas density $\rho_{\rm gas}$ decrease $\alpha_3$, making the \ac{GC} more top-heavy (more high-mass stars per star formed).
We note that this observationally motivated \ac{IMF} variation (Eqs. \ref{eq_alpha1} to \ref{eq_y}) needs to be further tested, but the variation fulfils the observational constraints from resolved stellar populations and the Galactic field as detailed in \cite{2013pss5.book..115K,2018A&A...620A..39J,2021A&A...655A..19Y}.
The deduced top-heavy \ac{IMF} for massive \acp{GC} changes the dynamical behaviour of the clusters compared to the canonical \ac{IMF} \citep{2020ApJ...904...43H,2021MNRAS.502.5185M,2021A&A...655A..71W}.

As in Paper I, the IMF was used to compute the mean stellar mass:
\begin{align}
    \left< m \right> = \frac{\int\limits_{m_{\rm low}}^{m_{\rm up}} dm~m \xi (m)}{\int\limits_{m_{\rm low}}^{m_{\rm up}} dm ~ \xi (m)},
\end{align}
where $m_{\rm low} = 0.08 M_\odot$ and $m_{\rm up}$ are the minimum and maximum stellar masses in the \acp{GC}, respectively.
$m_{\rm up}$ is computed for each \ac{GC} by solving eqs. 2 - 4 of \cite{2017A&A...607A.126Y}, replacing the fixed value of $m_{\rm up} = 120 ~M_\odot$ used in Paper I.

\subsubsection{Mass lost through stellar evolution}
\label{sec_MLstelEvo}

\begin{figure}
    \includegraphics[scale = 0.85]{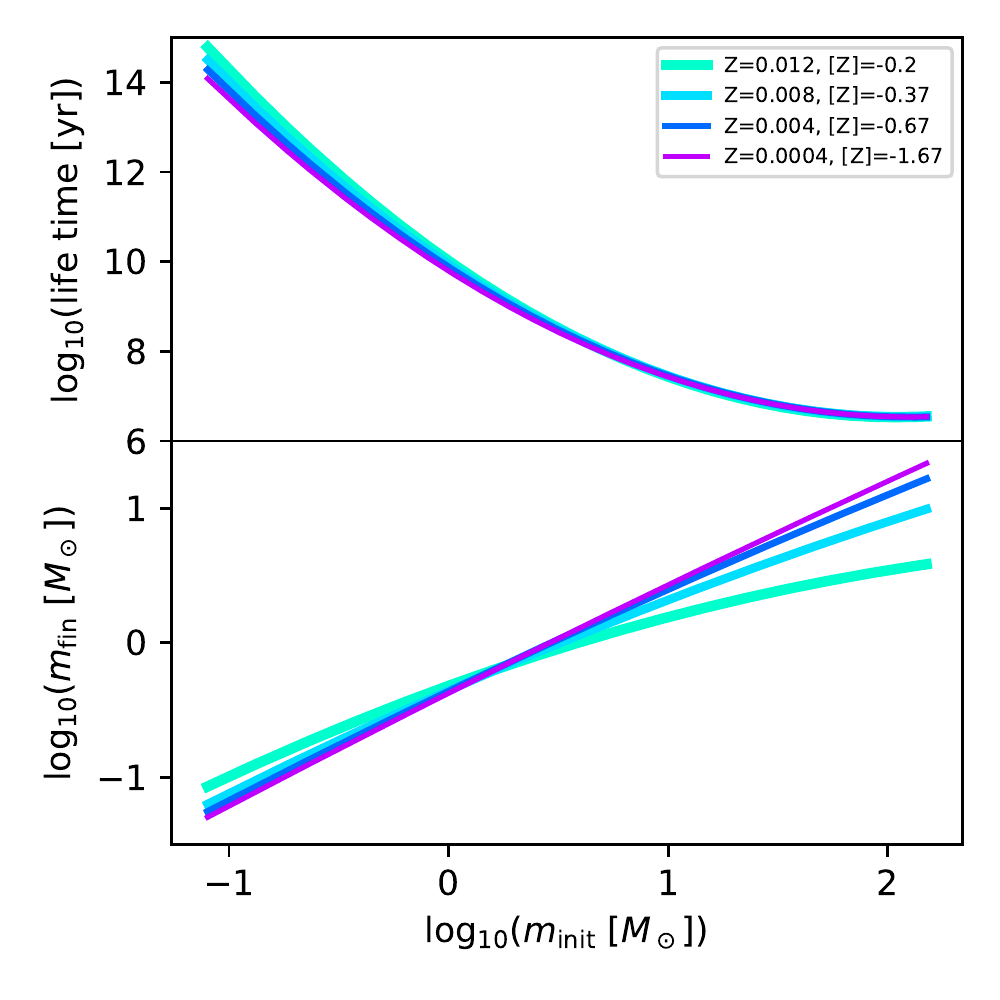}
    \caption{The stellar lifetimes and remnant masses, $m_{\rm fin}$, of stars as a function of their initial masses, $m_{\rm init}$, adapted from fig. 3 of \citet{2019A&A...629A..93Y}.}
    \label{fig_stellarLife}
\end{figure}

\begin{figure}
    \includegraphics{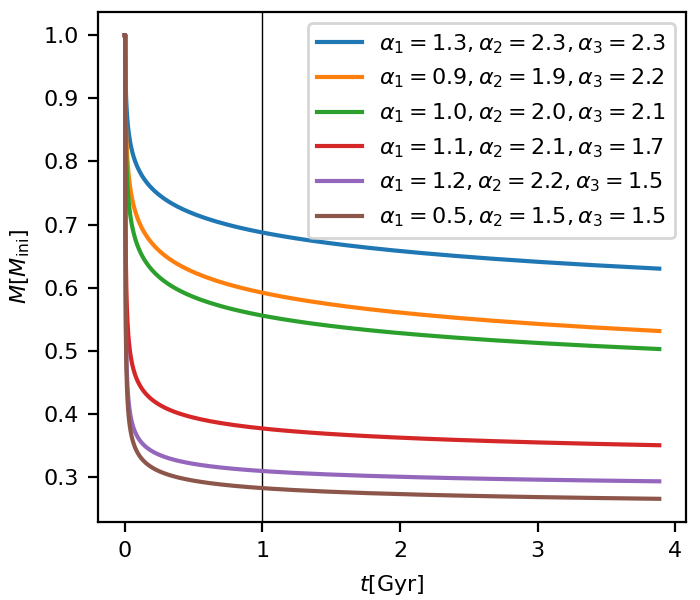}
    \caption{The remaining cluster mass, as mass fraction of the initial mass, taking only stellar evolution into account for different IMFs.
    t = 1 Gyr is marked with the vertical thin black line.
    The canonical IMF is given by the blue line.
    The other IMFs assume from top to bottom: $\left(\frac{\rho_{\rm gas}}{10^6 M_\odot pc^{-3}}, [Z/H]_{\rm ini}\right) =$ $(0.21, -0.55)$, $(0.38, -0.48)$, $(3.76, -0.41)$, $(11.93, -0.35)$, $(8.83, -1.28)$.}
    \label{fig_stellarEvoMassLoss}
\end{figure}

To compute $p_{\rm SF}$, the life expectancies and remnant masses for stars given in Fig. \ref{fig_stellarLife} are used.
These were obtained by \cite{2019A&A...629A..93Y} by fitting splines (a function defined piecewise as polynomials) to data provided in \cite{1998A&A...334..505P} and \cite{2001A&A...370..194M}.
These authors obtained their data from theoretical models.
To find $p_{\rm SF}$ the fraction of the initial mass of a \ac{GC} that is left depending on its age needs to be computed, neglecting dynamical evolution and thus only focussing on the effects of stellar evolution.

Computing the mass loss of a star over time requires complicated numerical models \citep{2016ApJ...821...38S,2022MNRAS.tmp..580F}.
However, stars loose most of their mass, after hydrogen burning is over i.e. towards the end of their life \citep{2000ARA&A..38..573W,2003ApJ...591..288H}.
This justifies the simplification used in the present work that all of the mass a star looses is subtracted at the end of its life.
It follows then that in a given time interval a \ac{GC} looses the difference between the initial mass and the remnant mass of all stars dying within this time.
The \ac{IMF} gives us the mass distribution of stars in the \ac{GC}, with which we can now compute the mass portion of $M_{\rm ini}$ left over in a \ac{GC} after accounting for stellar evolution at any given time.

Fig. \ref{fig_stellarEvoMassLoss} shows the evolution of the \ac{GC} mass computed as above for various combinations of $\alpha_1$, $\alpha_2$ and $\alpha_3$.
As we can see, the majority of a \ac{GC}'s mass loss through stellar evolution happens before the 1 Gyr mark, while the impact made by dynamical evolution is expected to still be small at this point given the long two body relaxation times.
This is no hard boundary and one could easily use $0.5$ or $1.5 ~\rm Gyr$ instead.
It is worth noting that core collapse happens quicker for \acp{GC} with more top-heavy \ac{IMF} and for \acp{GC} with a higher central density.
This might be taken into account in future models \citep{2004ApJ...604..632G}.
The dominance of the mass loss through stellar evolution is consistent with the findings of \citet[][]{2003MNRAS.340..227B} (see their fig. 1), while \citet{2001ApJ...550..691J} and \citet{2010MNRAS.409..305L} find slightly higher values for the time for which mass loss through stellar evolution dominates (up to $10 ~\rm Gyr$).
However, as Fig. \ref{fig_stellarEvoMassLoss} shows, the additional mass loss after 1 Gyr is small.
The mass of stars dying at this time is around $2 ~M_\odot$ which classify as cool stars.
Therefore, this time is used to estimate $p_{\rm SF}$.

As mentioned below Eq. \ref{eq_Mini}, the mass loss through stellar evolution is assumed to happen instantaneously at the beginning of \ac{GC} evolution.
This is possible since this mass loss dominates with respect to the dynamical mass loss during the \ac{GC}'s initial evolution, but plays only a small role afterwards \citep{2003MNRAS.340..227B}.
A more detailed numerical study of \acp{GC} and their mass loss depending on the \ac{IMF} can be found in \cite{2020ApJ...904...43H}.

It is worth noting that the mass lost through stellar evolution from the initial population is not added to the remaining gas mass from which the enriched stars form.
In the present sample, the mass of the gas returned would have been between $0.1$ and $0.2 ~M_{\rm ini}$ for most \acp{GC}.
The \ac{GC} experiencing the largest gas return is NGC 6441, which would regain 41 \% of its initial mass in gas.
This would add up to an extra 18 \% more gas to the amount of gas to be polluted in Eq. \ref{eq_Miron}.
Therefore, gas return can be neglected.

\subsubsection{The parameters $\beta$ and $x$}
\label{sec_metBetaX}

\cite{2003MNRAS.340..227B} fitted $\beta$ and $x$ to their Nbody results only for clusters with a King concentration parameter $W_0 = 5.0$ and $W_0 = 7.0$ with $x = 0.75, \beta = 1.91$ and $x = 0.82, \beta = 1.03$, respectively.
However, in this work, $\beta$ and $x$ for $W_0$ between 0.7 and 9 is needed, which is the range of $W_0$ we find for the \acp{GC} in our sample.
Therefore, $W_0$ is computed depending on the initial mass and then extrapolated linearly from the values given above.

To this end the tidal radius, $r_t$, is calculated as given by eq. 1 of \citet{2003MNRAS.340..227B},
\begin{align}
    \label{eq_tidalR}
    r_t = \left( \frac{G M_{\rm ini}}{2 V_{\rm G}^2} \right)^\frac{1}{3} R_{\rm p}^\frac{2}{3},
\end{align}
where G is the Newtonian gravitational constant, $V_G$ is the circular velocity of the Galaxy at the coordinates of the \ac{GC} and $R_{\rm p}$ the pericentre distance of the \ac{GC}.
As in Paper I,  $V_{\rm G}$ is assumed to be ${220 ~\rm km ~s^{-1}}$.
Additionally, the King radius, $r_0$, is needed.
It is defined in eq. 8.76 of \citet{2008LNP...760.....A},
\begin{align}
    r_0 &= \left( \frac{9}{4\pi G} \frac{\sigma^2}{\rho} \right)^{\frac{1}{2}},\\
    \rlap{with}\notag\\
    \sigma &= \sqrt{\frac{G M_{\rm ini}}{\epsilon r_{\rm h}}}, &\rho &= \frac{3 M_{\rm ini}}{8 \pi r_{\rm h}^3}, &r_h &= 0.10 {\rm pc} \left( \frac{M_{\rm ini}}{M_\odot} \right)^{0.13}.\notag
\end{align}
The definition of $r_h$ is taken from \cite{2012A&A...543A...8M}.
With this the concentration parameter defined as eq. 8.95 in \citet{2008LNP...760.....A} can be computed,
\begin{align}
    c = \log_{10}\left( \frac{r_t}{r_0} \right).
\end{align}
Fig. 8.5 of \cite{2008LNP...760.....A} shows the correlation between $c$ and $W_0$.
Fitting a linear curve going through the graph's origin to the data using the nonlinear least-squares Marquardt-Levenberg algorithm function \citep[see e.g.][]{doi:10.1137/0111030}, the following correlation is obtained:
\begin{align}
    W_0(c) &= (4.38 \pm 0.02) c.
\end{align}
Interpolating $\beta$ and $x$ linearly as mentioned at the beginning of this section leads to:
\begin{align}
    \beta &= 4.11 - 0.44 W_0,\\
    x &= 0.575 + 0.035 W_0.
\end{align}

With this all quantities required by Eq. \ref{eq_Mini} can be expressed in terms of $M_{\rm ini}$.
As in Paper I we use the Newton-Raphson method to solve Eq. \ref{eq_Mini}.

\subsubsection{The resulting numbers of SNe}
\label{sec_SNe}

The number of \acp{SN}, $N_{\rm SN}$, needed to produce the observed iron spread is computed analogous to Paper I:
\begin{align}
    N_{\rm SN} =& \frac{M_{\rm iron}}{0.074 M_\odot},\\
    \label{eq_Miron}
    \begin{split}M_{\rm iron} =& ~Z_{\rm Fe,\odot} \left( 10^{[Fe/H] + \sigma_{[Fe/H]}} - 10^{[Fe/H] - \sigma_{[Fe/H]}} \right) \\
         &\times M_{\rm ecl} \left( \frac{1}{\epsilon} - 1 \right),
         \end{split}
\end{align}
with the mean iron abundance of the \ac{GC}, $[Fe/H]$, the iron abundance spread, $\sigma_{\rm [Fe/H]}$, and the fraction of iron in the Sun, $Z_{\rm Fe,\odot} = 0.0013$ \citep{2009ARA&A..47..481A}.
The value of $0.074 M_\odot$ is the average mass of iron released by a \ac{SN} as computed by \citet{2017ApJ...848...25M}.

\subsection{The time after which star formation ends}

To compute the time after which \ac{SF} ends, the equations from Paper I need to be adjusted.
Firstly, the IMF needs to be integrated to find the number of expected \ac{SN}, $N_{\rm SN}^{\rm exp}$, for each \ac{GC} separately, instead of having a fixed fraction of the total number of stars as in Paper I:
\begin{align}
    N_{\rm SN}^{\rm exp} = \int\limits_{m_{\rm SN}}^{m_{\rm up}} dm ~\xi(m),
\end{align}
where $m_{SN} = 8 M_\odot$ is the mass above which stars end their lives as \acp{SN}.
As explained in Sec. \ref{sec_SNe}, $N_{\rm SN}$ of these \acp{SN} are required to explain the observed iron spread and, therefore, $N_{\rm SN}$ is the number of \acp{SN} that contribute to \ac{SF}.
The most massive stars are expected to explode first \citep[Fig. \ref{fig_stellarLife},][]{2019A&A...629A..93Y}.
This means that the mass, $m_{\rm last}$, of the last star to explode in a \ac{SN} which still contributes to \ac{SF} can be infered from $N_{\rm SN}$ and $N_{\rm SN}^{\rm exp}$:
\begin{align}
    \frac{N_{\rm SN}}{N_{\rm SN}^{\rm exp}} &= \frac{\int\limits_{m_{\rm last}}^{m_{\rm up}} dm ~\xi(m)}{\int\limits_{m_{\rm SN}}^{m_{\rm up}} dm ~\xi(m)} = \frac{ \frac{k_3}{1-\alpha_3} \left(m_{\rm up}^{1-\alpha_3} - m_{\rm last}^{1-\alpha_3} \right)}{\int\limits_{m_{\rm SN}}^{m_{\rm up}} dm ~\xi(m)},\\
    m_{\rm last} &= \left( m_{\rm up}^{1-\alpha_3} - \frac{N_{\rm SN} }{N_{\rm SN}^{\rm exp}} \frac{1 - \alpha_3}{k_3} \int\limits_{m_{\rm SN}}^{m_{\rm up}} dm ~\xi(m) \right)^{\frac{1}{1-\alpha_3}}.
\end{align}
Since $m_{\rm last}$ is the mass of a star that explodes as a \ac{SN}, it has to be between the minimum mass from which a star becomes a \ac{SN}, $8 M_\odot$, and the maximum stellar mass, computed as explained in Sec. \ref{sec_metClM_IMF}.
As in Paper I, the expected lifetime for a star with mass $m_{\rm last}$ can be looked up in Fig. \ref{fig_stellarLife} for $[Fe/H] = -1.67$, which is close to the $[Fe/H]$-value of most of the \acp{GC} in the present sample (Tab. \ref{tab_GCProps}).
This lifetime is used as the estimate for the time when further \ac{SF} ends.

\section{Results}
\label{sec_res}

\begin{table*} $~$
    \caption{The observed properties and deduced numbers of CCSNe and times when SF ends for 55 Galactic GCs.
    The columns from left to right are: the name of the GC, the current mass, $M(t)$, with $t=12 ~\rm Gyr$, the pericentre of the \ac{GC}'s orbit, $R_{\rm p}$, the apocentre of the \ac{GC}'s orbit, $R_{\rm ap}$, the computed value for $\alpha_3$, initial mass, $M_{\rm ini}$, iron abundance, $[Fe/H]$, iron abundance spread, $\sigma_{[Fe/H]}$, number of CCSNe, $N_{\rm SN}$, the  number of CCSNe per unit mass, $n_{\rm SN}$, the expected number of CCSNe to occur in the GC, $N_{\rm SN}^{\rm exp}$, and the time SF ends, $t_{\rm end}^{\rm SF}$.}
    \input{tables/LatexTable1.tex}
    \label{tab_GCProps}
\end{table*}

The results calculated for a top-heavy \ac{IMF} are compiled in Tab. \ref{tab_GCProps}.
As in Paper I the metallicities cited were taken from \cite{2019ApJS..245....5B} and the present day \ac{GC} masses and orbital parameters are taken from \cite{hilker_baumgardt_sollima_bellini_2019}.
Note that an upgraded catalogue for the metallicities is available \citep{2022ApJ...925...36B}, however, for consistency with Paper I the old version \citep{2019ApJS..245....5B} is used for this work.
The data in Table \ref{tab_GCProps} show that the obtained values for the mass function index, $\alpha_3$, are all smaller than $2.3$ and, therefore, the \ac{IMF} is more top-heavy than the canonical \ac{IMF} for all of the \acp{GC}.
In the following sections the individual properties and how they compare to the results in Paper I will be discussed.
The results from Paper I are marked with a superscript `${\rm can}$' (to stand for canonical).

\subsection{The initial masses}
\label{sec_ResMini}

\begin{figure}
    \includegraphics{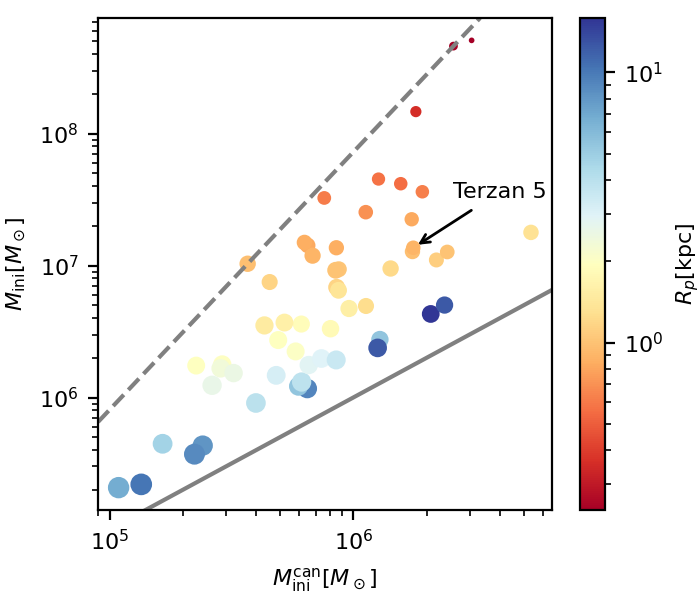}
    \caption{The initial masses computed for the GCs compared to the results from Paper I.
    The identity function has been marked with a solid grey line and $\frac{M_{\rm ini}}{M_\odot} = 1.4 \times 10^{-4} \left( \frac{M_{\rm ini}^{\rm can}}{M_\odot} \right)^{2.0}$ has been marked with a dashed grey line.
    The sizes of the dots correlate linearly with $\alpha_3$ (smaller means a more top-heavy IMF) and the colour indicates the pericentre distance $R_{\rm p}$.
    The cluster Terzan 5 (Sec. \ref{sec_ResMini}, \ref{sec_TimeSF}) is marked.}
    \label{fig_MiniOldMini}
\end{figure}

\begin{figure}
    \includegraphics{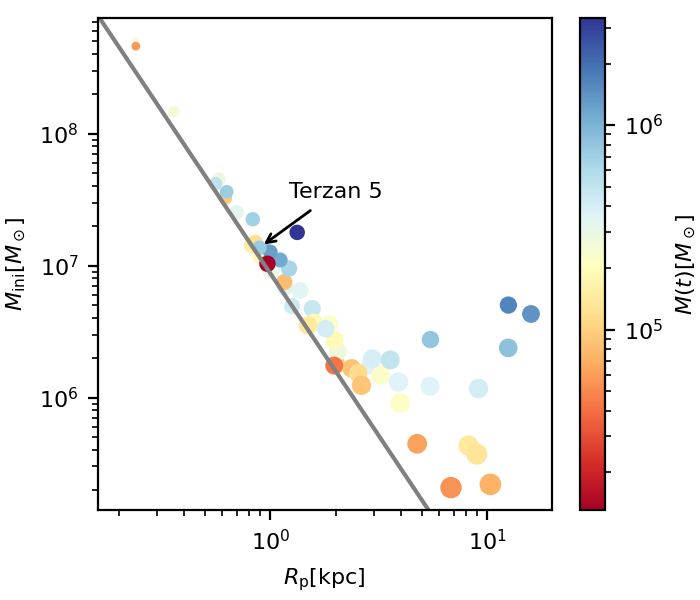}
    \caption{The initial masses computed for the GCs in dependence of the pericentre distance.
    Eq. \ref{eq_RpMini} has been marked with a grey line.
    The sizes of the dots correlate linearly with $\alpha_3$ (smaller means a more top-heavy IMF) and the colour indicates the present-day mass $M(t)$.
    The cluster Terzan 5 (Sec. \ref{sec_ResMini}, \ref{sec_TimeSF}) is marked.}
    \label{fig_RpMini}
\end{figure}

\begin{figure}
    \includegraphics{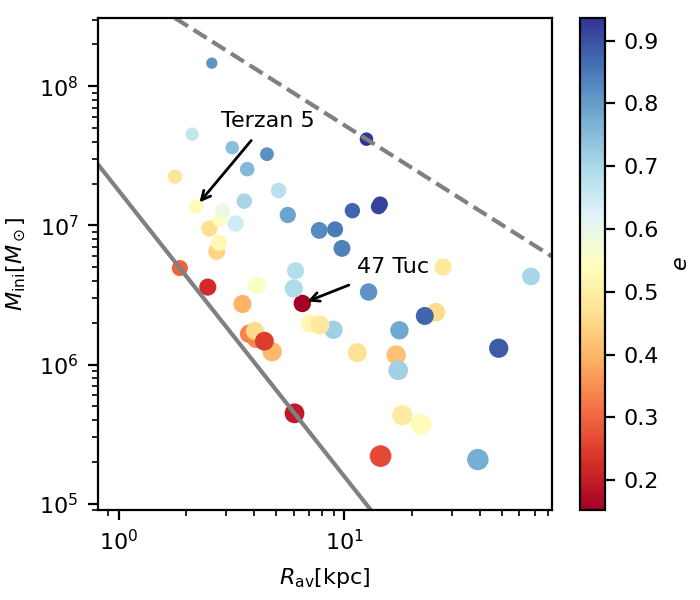}
    \caption{The initial masses computed for the GCs in dependence of the mean orbital distance $R_{\rm av}$.
    The function $\frac{M_{\rm ini}}{M_\odot} = 1.8 \times 10^{7} \left(\frac{R_{\rm av}}{\rm kpc}\right)^{-2.1}$ has been marked with a solid grey line and $\frac{M_{\rm ini}}{M_\odot} = 5.6 \times 10^{8} \left(\frac{R_{\rm av}}{\rm kpc}\right)^{-1.0}$ has been marked with a dashed grey line.
    The sizes of the dots correlate linearly with $\alpha_3$ (smaller means a more top-heavy IMF) and the colour indicates the orbital eccentricity $e$.
    The clusters Terzan 5 (Sec. \ref{sec_ResMini}, \ref{sec_TimeSF}) and 47 Tuc (Sec. \ref{sec_ConsGal}) are marked.}
    \label{fig_RavMini}
\end{figure}

\begin{figure}
    \includegraphics{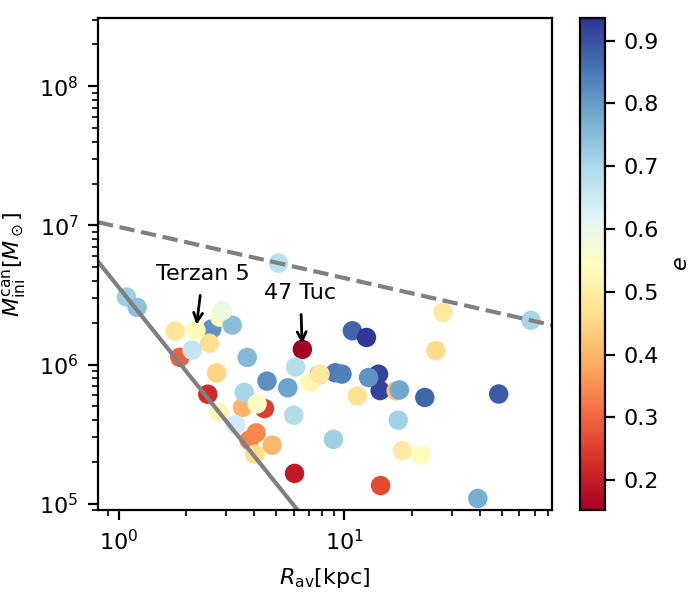}
    \caption{The same as Fig. \ref{fig_RavMini} but the initial masses are computed using the canonical IMF.
    The function $\frac{M_{\rm ini}^{\rm can}}{M_\odot} = 3.6 \times 10^{6} \left(\frac{R_{\rm av}}{\rm kpc}\right)^{-2.0}$ has been marked with a solid grey line and $\frac{M_{\rm ini}^{\rm can}}{M_\odot} = 9.8 \times 10^{6} \left(\frac{R_{\rm av}}{\rm kpc}\right)^{-0.37}$ has been marked with a dashed grey line.}
    \label{fig_RavMiniOld}
\end{figure}

In Figs. \ref{fig_MiniOldMini} to \ref{fig_NsnOldNsn} limiting functions were fitted to the data.
The method for obtaining these functions is described in Appendix \ref{app_Comp}.
The code to fit these functions can be found on the authors' github repository\footnote{\url{https://github.com/Henri-astro/LimiFun}}.

In Fig. \ref{fig_MiniOldMini} the new results computed using a variable \ac{IMF}, $M_{\rm ini}$, are compared with the $M_{\rm ini}^{\rm can}$ computed using the canonical \ac{IMF} (Paper I).
The differences between $M_{\rm ini}$ and $M_{\rm ini}^{\rm can}$ sensitively depend on the assumed IMF variation law, that is, Eqs. \ref{eq_alpha1} to \ref{eq_y}.
For small $\alpha_3$ the increase of $M_{\rm ini}$ compared to $M_{\rm ini}^{\rm can}$ is larger.
That means that \acp{GC} with more top-heavy \acp{IMF} loose mass more rapidly than \acp{GC} with larger $\alpha_3$.
This is in agreement with \cite{2019MNRAS.482.5138B}\footnote{Note that in \cite{2019MNRAS.482.5138B} the IMF power-law indices, $\alpha_i$, are defined with the opposite sign.}.
However, while \cite{2019MNRAS.482.5138B} conclude that the combined mass of all early \acp{GC} in the Galaxy must have been some $10^8 M_\odot$, according to the present study calculations the 55 \acp{GC} in our sample already exceed this mass.
Part of the reason is the higher maximum stellar mass in this work.
\cite{2019MNRAS.482.5138B} used only $m_{\rm up} = 15 M_\odot$ for all of their \acp{GC}, while in this study the $M_{\rm ecl}$-$m_{\rm up}$-relation found by \cite{2017A&A...607A.126Y} is used.
With this, values for $m_{\rm up}$ between 140 and $150 ~\rm M_\odot$ are found for the \acp{GC} studied here.
Consequently, in the most massive \acp{GC} studied here up to 80 \% of the initial mass is in stars more massive than $15 M_\odot$.

Another question the top-heavy \ac{IMF} leads to is, how many stars can form in a \ac{GC} before star formation is suppressed due to the heating of the gas.
Several studies showed that stellar winds and radiation can disperse the gas in a stellar cluster before the first \acp{SN} occur \citep{2003ARA&A..41...57L,2012MNRAS.424..377D,2022arXiv220202237V}.
The overabundance of massive stars relative to the canonical \ac{IMF} would increase these effects and could eventually prevent iron-enhanced stars from forming.
However, further investigation is required to estimate the exact magnitude of these effects depending on the \ac{IMF} and the birth gas density, $\rho_{\rm gas}$ and $[Fe/H]$.

As expected from Eqs. \ref{eq_alpha3} and \ref{eq_rho} and visible in Fig. \ref{fig_MiniOldMini}, $M_{\rm ini}$ increases with decreasing $\alpha_3$.
Since $\alpha_3$ is below the value for the canonical \ac{IMF} for all \acp{GC}, $M_{\rm ini}$ is always larger than $M_{\rm ini}^{\rm can}$.
The difference between $M_{\rm ini}$ and $M_{\rm ini}^{\rm can}$ increases with decreasing $R_{\rm p}$.
This is partly because the tidal radius computed using Eq. \ref{eq_tidalR} is the tidal radius of the \ac{GC} at its pericentre.
This means that the mass loss for low-$R_{\rm p}$-\acp{GC} is stronger, which means that their initial masses were larger and therefore their $\alpha_3$ is increased.
A similar but weaker correlation can be found, if a similar plot is done with $R_{\rm ap}$, as defined below Eq. \ref{eq_Mini}, instead of $R_{\rm p}$.
This leads to the conclusion that the difference between $M_{\rm ini}$ based on a top-heavy \ac{IMF} compared to $M_{\rm ini}^{\rm can}$ based on a canonical \ac{IMF} is larger for \acp{GC} closer to the Galactic centre.

The smaller $R_{\rm p}$ is for a \ac{GC}, the stronger is the tidal force from the Galactic gravitational field it is exposed to.
This means that the dissolution times for \acp{GC} get smaller with decreasing $R_{\rm p}$.
Therefore, if a \ac{GC} comes close enough to the Galactic centre, its dissolution time becomes smaller than the age of the present day \acp{GC}, ${12~\rm Gyr}$ \citep{2010ApJ...708..698D,2019MNRAS.490..491U,2021arXiv210908708C}.
As visible in Figs. \ref{fig_MiniOldMini} and \ref{fig_RpMini}, there is a lack of low-$R_p$-\acp{GC} at the lower end of $M_{\rm ini}$.
The lower limit for not yet dissolved \acp{GC} is found at:
\begin{align}
\label{eq_RpMini}
\mathrm{\frac{M_{\rm ini}}{M_\odot} = 8.7 \times 10^{6} \left(\frac{R_{\rm p}}{\rm kpc}\right)^{-2.5}.}
\end{align}
That this is the lower limit for undissolved \acp{GC} is further supported by the fact that the present-day masses of the \acp{GC} get smaller for \acp{GC} closer to the boundary.
It is, however, likely for these \acp{GC} to still exist as dark clusters to this day \citep{2011ApJ...741L..12B}.
Furthermore, there appears to be a lack of distant initially massive \acp{GC}.
This is even more visible in Fig. \ref{fig_RavMini}, which shows $M_{\rm ini}$ over the mean orbital distance over time, $R_{\rm av} = \frac{R_{\rm ap} + R_{\rm p}}{2}\left( 1 + \frac{e^2}{2} \right)$ \citep{10.2307/2689506}.
This would suggest that more massive \acp{GC} formed preferably near the Galactic centre (small $R_{\rm av}$).
In Fig. \ref{fig_RavMiniOld} it is shown that this result holds even if the canonical \ac{IMF} is used, which indicates that this result is robust.

It is important to note that in both this work and in \cite{2003MNRAS.340..227B}, $R_p$ was used to compute the tidal radius (see Sec. \ref{sec_metBetaX}).
$R_p$ is the smallest distance the \ac{GC} can have from the Galactic centre.
This leads to an underestimate of the tidal radius and therefore an overestimate of the mass loss of a \ac{GC}.
This overestimation is larger for larger eccentricities, since the orbit of a \ac{GC} with large eccentricity deviates more from a circular orbit with radius $R_{\rm p}$.
As visible in Fig. \ref{fig_RavMini} the \acp{GC} close to the lower boundary (lower grey line) all have very low eccentricities, therefore, the $M_{\rm ini}$ values are expected to be close to the real ones.
However, the \acp{GC} close to the upper boundary (upper grey line) all have high eccentricities, indicating that we are likely to have overestimated their masses.
To determine the actual extent of this overestimate, the simulations of \cite{2003MNRAS.340..227B} would have to be rerun with more precise estimates of the tidal radius.
47 Tuc is an outlier when it comes to the eccentricity and has been labelled in both Figs. \ref{fig_RavMini} and \ref{fig_RavMiniOld}.

\subsection{The number of SNe and the time SF ends}
\label{sec_TimeSF}

\begin{figure}
    \includegraphics{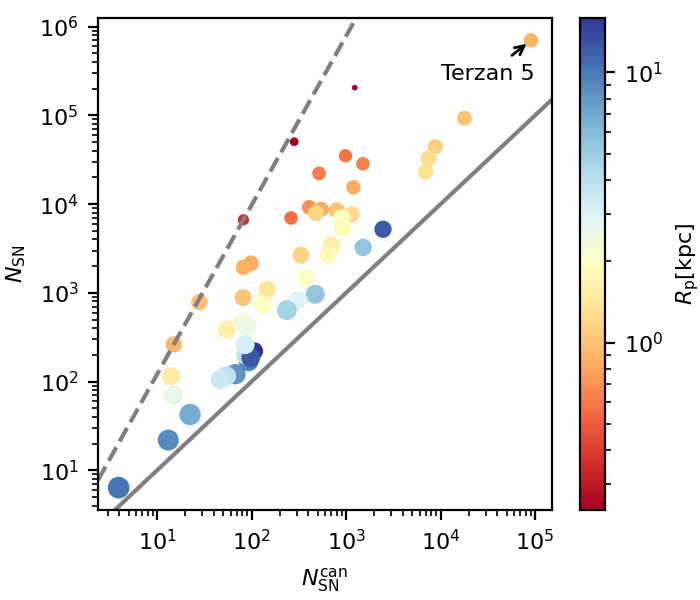}
    \caption{The required numbers of SN computed for the GCs compared to the results from Paper I.
    The identity function has been marked with a solid grey line and $N_{\rm SN} = 1.5 \left({N_{\rm SN}^{\rm can}}\right)^{1.9}$ has been marked with a dashed grey line.
    The size of the dots correlates linearly with $\alpha_3$ (smaller means a more top-heavy IMF) and the colour indicates the Galactocentric pericentre distance.
    The cluster Terzan 5 (Sec. \ref{sec_ResMini}, \ref{sec_TimeSF}) is marked.}
    \label{fig_NsnOldNsn}
\end{figure}

One of the key results of Paper I was that the number of \acp{SN} per unit mass required to reproduce the observed metal-enrichment, $n_{\rm SN}$, is independent of $M_{\rm ini}$.
Therefore, $n_{\sc SN}$ does not change between this work and Paper I (compare Tab. \ref{tab_GCProps} and tab. 1 in Paper I).
However, the change in $M_{\rm ini}$ propagates through to $N_{\rm SN}$ in the form $N_{\rm SN} = n_{\rm SN} M_{\rm ini}$.
This leads to the behaviour visible in Fig. \ref{fig_NsnOldNsn}, which is similar to that for $M_{\rm ini}$ in Fig. \ref{fig_MiniOldMini}.
The difference between $N_{\rm SN}$ and $N_{\rm SN}^{\rm can}$ is larger for \acp{GC} closer to the Galactic centre.

As in Paper I, the number of \acp{SN} required to produce the iron spread observed in Terzan 5 is larger than $N^{\rm exp}_{\rm SN}$.
This leads to the conclusion that Terzan 5 must have been formed through a different scenario, like for example a merger \citep[Paper I]{2014ApJ...795...22M}.
\acp{GC} with unusual chemical compositions can also form through the merger of two chemically distinct molecular clouds \citep{2022arXiv220705745H}.
Terzan 5 has been marked in all relevant figures.

\begin{figure}
    \includegraphics{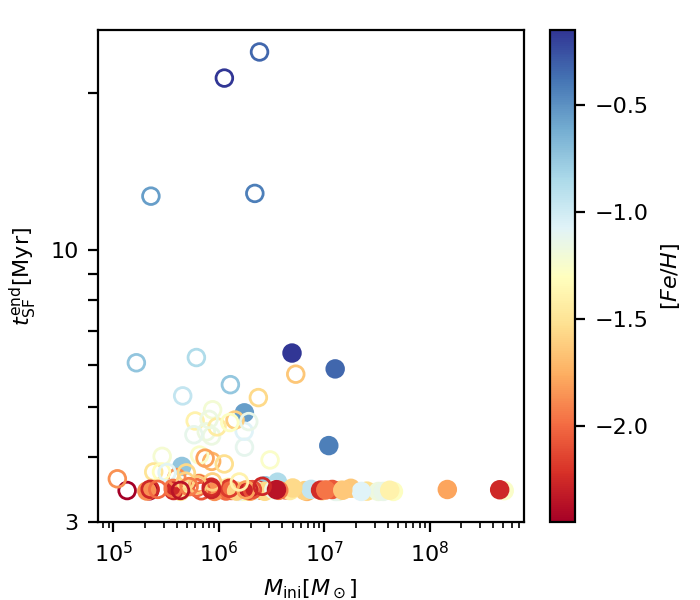}
    \caption{The time until SF ends. The empty circles are the datapoints computed using the canonical IMF from Paper I, the filled ones are from this work.
    The colour indicates the iron abundance.}
    \label{fig_Tsf}
\end{figure}

\begin{figure}
    \includegraphics{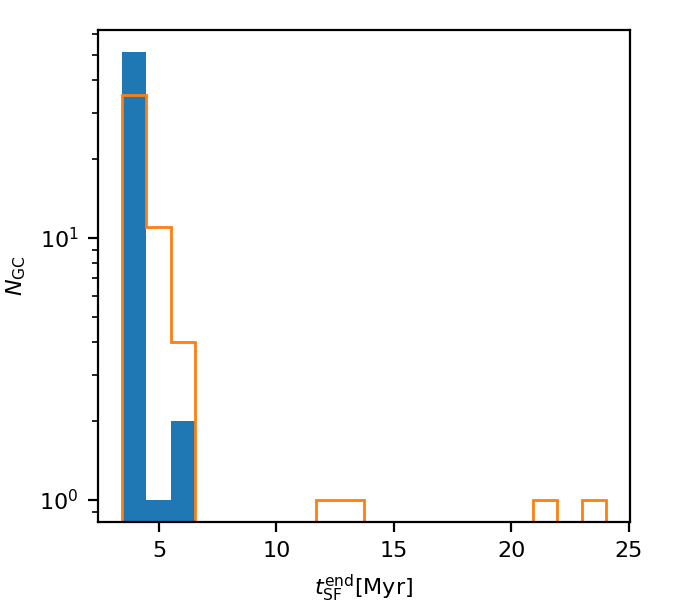}
    \caption{Distribution of the times, $t_{\rm SF}^{\rm end}$, when star formation ends for the clusters in our sample.
    $N_{\rm GC}$ stands for the number of clusters in a given bin.
    The blue bars show the data computed in the present work, the orange bars show the data computed using a canonical IMF in Paper I.
    Note that the values of $t_{\rm SF}^{\rm end}$ have a smaller dispersion for the variable IMF.}
    \label{fig_HistTsf}
\end{figure}

In Fig. \ref{fig_Tsf} the time when \ac{SF} ends is plotted over $M_{\rm ini}$ and compared to our data from Paper I.
Note that Terzan 5 is missing, since the required number of \acp{SN} exceeds the number of \acp{SN} expected for this \ac{GC} (see Tab. \ref{tab_GCProps}).
This makes it impossible to compute $t_{\rm SF}$.
The calculated times until SF ends depend on the assumed lifetime of massive stars (Fig. \ref{fig_stellarLife}).
Different stellar evolution models can result in different lifetimes with uncertainties of a few Myr.
The increase of $M_{\rm ini}$ discussed in Sec. \ref{sec_ResMini} is clearly visible in this plot.
Additionally, the time until \ac{SF} ends is reduced significantly.
This is because \acp{GC} with top-heavy \acp{IMF} produce a larger fraction of their \acp{SN} early on due to the larger fraction of massive stars.
The times until \ac{SF} ends obtained here are in agreement with the times after which \acp{GC} have been observed to be free of gas \citep{2014MNRAS.443.3594B} and with theoretical models concerning the timescales with which gas is expelled from \acp{GC} via \acp{SN} \citep{2015ApJ...814L..14C}.

Similar to Fig. \ref{fig_Tsf}, Fig. \ref{fig_HistTsf} shows the time until \ac{SF} ends, however, this time in the form of a histogram.
It is clearly visible that $t_{\rm SF}^{\rm end}$ is shorter using a top-heavy \ac{IMF} compared to the results of our calculations using a canonical one.
For most \acp{GC}, \ac{SF} ends after ${3.5-4 ~\rm{Myr}}$.
As for the results computed with the canonical \ac{IMF}, this is in agreement with studies showing that the \acp{GC} and young massive clusters are gas free after at most $10 ~\rm Myr$ \citep{2015ApJ...814L..14C,2016A&A...587A..53K}.

The correlation between the mean iron abundance and the time until \ac{SF} ends observed in Paper I is also apparent in our data (Fig. \ref{fig_Tsf}) with a variable \ac{IMF}.
However, as in Paper I, this might be due to an underestimate of the error of the iron spread, which would increase the calculated number of \acp{SN} preferentially for \acp{GC} with large mean iron abundances.
Therefore, it is unclear whether or not this is a physical effect.
According to \cite{2012MNRAS.422.2246M}, a higher mean iron abundance is expected to lead to a less top-heavy \ac{IMF}.
Therefore, the portion of massive stars would be lower, which would lead to fewer \acp{SN} early on in \ac{GC} evolution.
This means that it is expected that gas is driven out more slowly leading to a larger $t_{\rm SF}^{\rm end}$, which would support the result we observe in Fig. \ref{fig_Tsf}.

\section{Discussion}
\label{sec_disc}

\subsection{The upper mass limit of \acp{GC}}
\label{sec_upMassLim}

As stated in Sec. \ref{sec_ResMini}, the upper initial mass limit of \acp{GC} decreases with increasing distance from the Galactic centre.
The same has been observed by \cite{2013MNRAS.435.2604P} for young star clusters around M33.
They suggested that this would be due to higher gas surface densities (more material) in the inner regions of galaxies.

The star clusters in M33 are very young \citep[$\leq 10 ~\rm Myr$,][]{2011A&A...534A..96S,2013MNRAS.435.2604P} and, therefore, the mass loss experienced by these clusters is expected to be small ($M_{\rm cl} \approx M_{\rm ini}$, with the current cluster mass $M_{\rm cl}$).
\cite{2013MNRAS.435.2604P} expressed the mass of the most massive clusters in their radially binned sample as a function of their current galactocentric distance, $R$, using the following expression:
\begin{align}
    \log_{10}\left(\frac{M_{\rm cl}}{M_\odot}\right) = a \frac{R}{\rm kpc} + b.
\end{align}
By fitting this function to the data from the clusters in M33, they found that the clusters with the highest mass roughly followed the function with $a = -1.7 \times 10^{-4}$ and $b = 4.67$.
To obtain their fit they distributed their clusters into bins of 17 clusters each depending on their radial distance to the centre of M33. They then used the $\frac{R}{\rm kpc}$- and $\log_{10}\left( \frac{M_{\rm cl}}{M_\odot}\right)$-values of the most massive clusters in each bin to obtain their fit.
In this work, the maximum \ac{GC} mass is fitted using the method explained in App. \ref{app_Comp} and which lead to $\log_{10}\left(\frac{M_{\rm ini}}{M_\odot}\right) = 8.7 - \log_{10}\left(\frac{R_{\rm av}}{\rm kpc}\right)$ as the upper boundary (see Fig. \ref{fig_RavMini}).
This means that the upper limits of the initial masses of the \acp{GC} in our Galaxy are about three orders of magnitude higher than the ones in M33.
Note, however, that in both cases the sample size is too small to reliably determine the shape of the radial upper limit function.

While both, M33 and the very young Galaxy have their newly-formed, most massive cluster masses decrease systematically with the galactocentric radius, the two cases are very different: M33 is a late-type disk galaxy about ten times less massive than the Galaxy \citep{2003MNRAS.342..199C,2011MNRAS.414.2446M,2012ApJ...761...98K,2017MNRAS.465...76M}, is on the galaxy main sequence and is forming open star clusters, while the GC system studied here consists of 12 Gyr old \acp{GC} in the Galactic halo.
Since the mean \ac{GC} mass increases with the mass of the host galaxy \citep{2013ApJ...772...82H} we expect differences in the maximum mass function for those two systems.

\subsection{Consequences for the Galaxy}
\label{sec_ConsGal}

Assuming the Galaxy and all \acp{GC} were also formed in-situ \citep{1998AJ....116..748G}, eqs. 9 - 12 from \cite{2017A&A...607A.126Y} can be used to compute the \ac{SFR} of the Galaxy at the time the \acp{GC} were forming.
With a maximum \ac{GC} mass of $5.1 \times 10^8 M_\odot$ for Terzan 1 we compute a \ac{SFR} of $351 M_\odot /\rm yr$.
This is about 100 times higher than the mean \ac{SFR} of the Galaxy and could be due to \ac{SFR} fluctuations, or due to the most massive \acp{GC} having a different origin.
It is, however, worth mentioning that especially for massive \acp{GC}, like $\omega$ Cen and Terzan 5, alternative formation scenarios, such as mergers or being the remnant of a tidally thrashed dwarf galaxy, have been proposed \citep[see e.g.][]{2003MNRAS.346L..11B,2014ApJ...795...22M,2022A&A...659A..96M}.
If the real maximum \ac{GC} mass was, for example, only half as large, the \ac{SFR} becomes $132 M_\odot /\rm yr$. If the three most massive \acp{GC} ($> 10^8 M_\odot$) of the sample are excluded, less than a tenth of the maximum \ac{GC} mass used above is needed.
The resulting \ac{SFR} is only $1 M_\odot /\rm yr$.

\begin{figure}
    \includegraphics{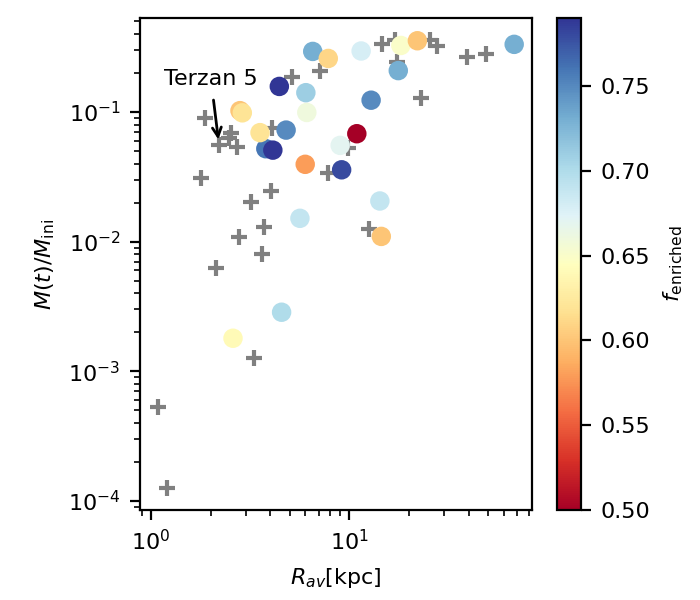}
    \caption{The fraction of mass left over in a GC at the current time over its Galactocentric distance.
    The GCs are colourcoded for the fraction of the enriched population within them.
    Those GCs for which the fraction of the enriched population is not known are marked with a `+'.
    The cluster Terzan 5 (Sec. \ref{sec_ResMini}, \ref{sec_TimeSF}) is marked.}
    \label{fig_RAvTMMini}
\end{figure}

As visible in Fig \ref{fig_RAvTMMini} the closer a \ac{GC} is to the Galactic centre, the larger the mass fraction it lost with respect to its initial mass.
This is a direct consequence of the results of the numerical experiments by \cite{2003MNRAS.340..227B}.
Contrary to this, \cite{2015MNRAS.453..357B} concluded, based on a lack of a connection between the radial distance of a \ac{GC} to the Galactic centre and the fraction of enriched stars within it, that no radial dependency between the cluster mass loss and its coordinates exists.
They expect a radial dependency to exist due to a preferential loss of 1st generation stars which are known to be less concentrated.
In a stronger tidal field they should, therefore, be lost faster and their present day fraction in the \ac{GC} should be lower.
\cite{2015MNRAS.453..357B} define first-generation stars as those with sodium abundance, $[Na/Fe]$, between $[Na/Fe]_{\rm min}$ and $[Na/Fe]_{\rm min} + 0.3$.
The minimum sodium abundance, $[Na/Fe]_{\rm min}$, of a \ac{GC} was determined by eye from $[O/Fe]$-$[Na/Fe]$-plots of the \acp{GC} excluding outlying stars.
However, \cite{2019MNRAS.487.3815M} find that in many \acp{GC} the difference between first and second generation is smaller than this. In fact, the $[Na/Fe]$-values of 1st and second generation stars overlap in some of the cases.

\cite{2015MNRAS.453..357B} used present-day fractions of enriched stars from a number of studies \citep{2009A&A...505..117C,2011A&A...533A..69C,2012ApJ...750L..14C,2013A&A...557A.138C,2013A&A...554A..81K,2014A&A...564A..60C,2015A&A...578A.116C,2015MNRAS.450..815M}.
Fig. \ref{fig_RAvTMMini} is colourcoded according to the data collected by \cite{2015MNRAS.453..357B}.
This figure confirms that there is indeed no obvious correlation between the fraction of enriched stars, $f_{\rm enriched}$, and the \ac{GC}'s distance from the Galactic centre or the mass fraction remaining.
However, due to the large number of datasets used, the unreliable boundary between different generations and the large scatter, this is not conclusive evidence against a preferential mass loss nearer to the Galactic centre.

The higher initial masses of the \acp{GC} would imply that a larger portion of the stars in the Galactic halo originated from \acp{GC}.
The mass of the stellar halo is currently estimated to be up to $1.4 \pm 0.4 \times 10^9 M_\odot$ \citep{2019MNRAS.490.3426D}.
Note, however, that this estimate was made using the canonical \ac{IMF} from \textit{Gaia} measurements of red giants in the halo.
A top-heavy galaxy-wide \ac{IMF} would increase the initial halo mass significantly.
Additionally, the number of stellar remnants would be increased \citep{2021A&A...655A..19Y}.
These remnants would either stay inside the \ac{GC} or be expelled due to dynamical processes, remaining as free-floates in the halo \citep{2008MNRAS.384.1231B}.
In our sample the number of neutron stars and black holes is on average 36 times higher than when assuming a canonical \ac{IMF} for all \acp{GC}.
The most extreme case in this study, Terzan 8, produces 403 times as many neutron stars and black holes as it would with a canonical \ac{IMF}.

\subsection{Improvements for future models}

Since a modified version of the method applied in Paper I is used, most of the limitations discussed in sec. 4.2 of Paper I apply in this work as well.
Among them is the possibility of failed \acp{SN} \citep{2015ApJ...801...90P,2016ApJ...821...38S,2020arXiv200715658B,2021arXiv210403318N}, in which case some massive stars would end their lives without contributing any iron.
This would lead to longer timespans until \ac{SF} ends, which is also discussed in \cite{2022arXiv220303002R}.

The dependence of the iron output on the \ac{SN} mass was also neglected.
However, it is likely, that the amount of ejected iron does depend on the mass of the progenitor.
From the short timespan after which \ac{SF} ends (Fig. \ref{fig_Tsf}), it follows that only the most massive stars contribute to the iron available for the formation of further stars.
Therefore, if for example, the most massive stars would contribute twice as much iron as the average \ac{SN}, only half as many \acp{SN} would be required and the time until \ac{SF} ends would become shorter.
Calculations by \cite{2013ARA&A..51..457N} show that, for example, a $40 M_\odot$-star with an initial metallicity of $[Z/H] = -1.7$ alone can produce $\approx 0.3 M_\odot$ in iron.
An important upgrade for our model in the future would therefore be to use more detailed \acp{SN} yield models to more accurately compute the number of \acp{SN} needed to explain the observed iron spreads.
Additionally, supernovae 1a can happen early as edge-lit type 1a SNe \citep{2003NewA....8..283R}.
This has also be neglected.

As shown in Paper I possible errors in measurements of $[Fe/H]$ have a significant influence on the result.
In appendix A of Paper I it is shown that an overestimate of the measurement errors for individual stars in the sample by 0.13 dex can lead to an overestimate of the \acp{SN} required by over 1 dex.
Similarly, errors in the mean $[Fe/H]$ for a \ac{GC}, which might occur as a result of over- or underestimating the sizes of the different stellar populations within the \acp{GC}, can lead to over and underestimates in the number of \acp{SN} required to explain the iron spread.
Since the iron spread is given in dex, it's value in absolute numbers depends on the value of $[Fe/H]$ (see Eq. \ref{eq_Miron}).
An underestimate of $[Fe/H]$ would lead to an underestimate of the number of \acp{SN} required and vice versa.
For example an underestimate of $[Fe/H]$ by 0.1 would lead to a 20\% underestimate of the number of \acp{SN} required.

One of the main assumptions of our model, not discussed thus far, is that the \acp{GC} move on constant Keplerian orbits.
However, various effects like encounters with other \acp{GC}, gas clouds and dynamical friction could have altered the orbits of the \acp{GC}.
If the \acp{GC} used to be further away then mass loss would also be reduced, which means that we have overestimated $M_{\rm ini}$.
The orbital evolution of \acp{GC} could be determined by `backwards-integrating' the paths of all \acp{GC} in the Galaxy as in \cite{2018CNSNS..61..160P}, though this would not cover interactions with \acp{GC} or molecular clouds that are already dissolved or unknown.
The model of \cite{2003MNRAS.340..227B} would have to be expanded to integrate over the orbital path rather than assuming a constant orbit.

Furthermore, a constant \ac{SFE} of 0.3 is assumed for each \ac{GC}.
In the current model a higher (lower) \ac{SFE} would lead to a lower (higher) fraction of the gas being left over, which means that less (more) iron was needed to be produced.
For a constant gas cloud mass before \ac{SF}, the mass of iron required, $M_{\rm iron}$, is proportional to $1 - \epsilon$.
Because of this, $N_{\rm SN}$ would increase for a smaller \ac{SFE}.
While the value of the \ac{SFE} is still unclear, \cite{2001AJ....122.1888A} suggested that the \ac{SFE} might vary with the density of the molecular cloud the \ac{GC} forms out of.
From Eq. \ref{eq_alpha3} we see that a higher density leads to a more top-heavy \ac{IMF}, which would lead to more \acp{SN} expected to explode early on, ensuring a self-regulation of the formation of a \ac{GC}.

Another pressing question to answer, is how to form new stars in a gas cloud already densely populated by stars, that heat the gas.
\cite{1999A&A...352..138P} suggested that the second generation of stars would form in the shock waves caused by \acp{SN}.
This would be in agreement with the present findings, however, it would also mean that the assumption of well-mixed gas in our model would need to be revisited.

\section{Conclusions}
\label{sec_concl}

In this work the initial masses and numbers of \acp{SN} required to explain the iron abundance spreads of 55 Galactic \acp{GC} is investigated.
Based on the assumption that the gas cloud a \ac{GC} forms out of is well mixed (has the same mean iron abundance at every point in space) and that it is polluted by \acp{SN} instantaneously, an analytical model is used.
The stars are assumed to form with the systematically variable \ac{IMF} in accordance with \cite{2021A&A...655A..19Y}, while the initial masses of the \acp{GC} are deduced considering stellar evolution as well as the dynamical mass loss of the star clusters, using the algorithm described in \cite{2003MNRAS.340..227B}.

The main results of this work summarize as follows:
\begin{enumerate}[wide, label=(\arabic*), labelwidth=!, labelindent=\parindent]
\item The initial masses of the \acp{GC} computed allowing for the metallicity and density dependent \ac{IMF} are larger than the ones calculated using the canonical \ac{IMF} (Paper I) by a factor of up to $\approx 10^2$.
This also increases the number of \acp{SN} required to explain the observed iron abundance spreads, as more iron needs to be produced to pollute the more massive gas cloud.
\item The computed mass function power-law index $\alpha_3$ for high-mass stars ($m > 1 M\odot$) is smaller than 2.3 for all \acp{GC}, which means that all \acp{GC} had \acp{IMF} that were more top-heavy than the canonical one.
This decreases the time until \ac{SF} ends since the fraction of massive stars is larger and therefore a larger portion of the \acp{SN} are from massive, short-lived stars \citep{2012ApJ...747...72D}.
According to our model, for the majority of Galactic \acp{GC} within the studied sample, the \ac{SF} ended after 3.5 to 4.0 Myr after their birth.
A larger portion of massive stars would also mean that the number of neutron stars and black holes created from the cluster stars is increased.
In the present study we find on average 36 times as many \acp{NS} and \acp{BH} being formed compared to the results assuming a canonical \ac{IMF} for all \acp{GC}.
The most extreme case in this study, Terzan 8, produces 403 times as many neutron stars and black holes than it would with a canonical \ac{IMF}.
\item Both the lower and the upper limit for the initial \ac{GC} mass is largest close to the Galactic centre and declines with the Galactocentric radius.
The first finding is easily explained as \acp{GC} near the Galactic centre dissolve quicker due to the stronger gravitational field.
The second finding points towards more massive \acp{GC} having formed closer to the Galactic centre than further away \citep[see also][]{2013MNRAS.435.2604P}.
To determine if this is a general trend for \ac{GC} masses within galaxies, more \ac{GC} systems need to be analysed.
This would give us valuable information to determine whether or not \acp{GC} were formed in-situ and thus during the earliest assembly phase of the Galaxy with implications for the formation of super-massive black holes \citep{2020MNRAS.498.5652K}.
The monolithic-collapse-model for the formation of galaxy spheroids, bulges and elliptical galaxies implies that the most massive clusters form in the innermost regions of the later galaxy.
These can appear as quasars \citep{2017A&A...608A..53J}.
In this model, the mass of the most-massive central star cluster correlates with the mass of the final galaxy.
While the formation of the galaxy occurs on the down-sizing time, the central star cluster's stellar black holes are compressed into a relativistic state by the in-falling gas.
They form, within a few hundred Myr, a super-massive black hole through gravitational-wave-emission-driven collapse of the central stellar-black-hole system \citep{2020MNRAS.498.5652K}.
\end{enumerate}

\section*{Acknowledgements}

We want to thank an anonymous referee for their useful comments.
The authors acknowledge support from the Grant Agency of the Czech Republic under grant number 20-21855S and through the DAAD-Eastern-Europe Exchange grant at Bonn University.
Z.Y. acknowledges support from the Fundamental Research Funds for the Central Universities under grant number 0201/14380049. Z.Y. acknowledges support through the Jiangsu Funding Program for Excellent Postdoctoral Talent under grant number 20220ZB54. Z.Y. acknowledges the support of the National Natural Science Foundation of China (NSFC) under grants No. 12041305 and 12173016. Z.Y. acknowledges the science research grants from the China Manned Space Project with NO.CMS-CSST-2021-A08 (IMF).

\section*{Data availability}

The data used here has been cited and is available in published form.



\bibliographystyle{mnras}
\bibliography{IronSpreadGC2} 

\appendix
\FloatBarrier

\section{Computing the limiting functions from point clouds}
\label{app_Comp}

\begin{figure}
    \includegraphics{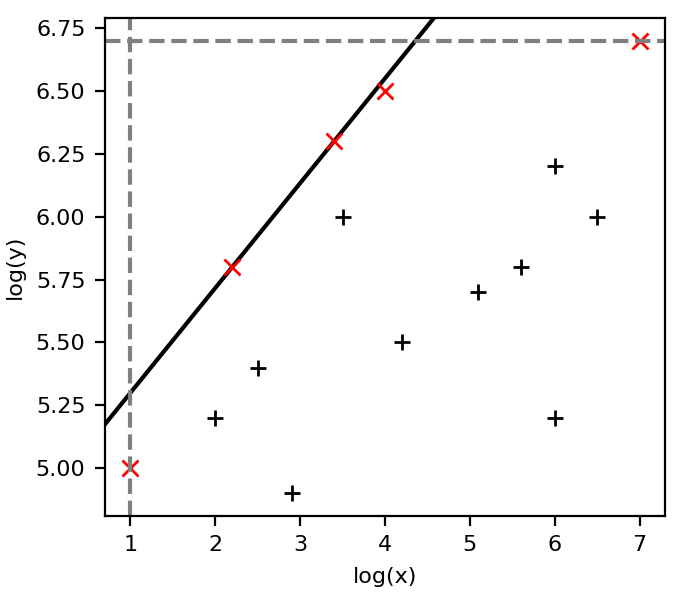}
    \caption{An example of a limiting function for the maximal values of a pointcloud given by the red and black crosses.}
    \label{fig_PointCloudEx}
\end{figure}

In this work the limiting functions to a number of point clouds has been computed (see Figs. \ref{fig_MiniOldMini} to \ref{fig_NsnOldNsn}).
All of these limiting functions are computed in log-log-space ($x' = \log(x)$ and $y' = \log(y)$) and under the assumption that the function takes the form $y' = a + bx'$ in this space.
The method will be demonstrated for the case of finding a maximum function with a positive slope to a point cloud as visible in Fig. \ref{fig_PointCloudEx} and then discuss how this algorithm can be generalized.
The algorithm attempts to find a line that satisfies two conditions:
\begin{enumerate}[wide, label=(\arabic*), labelwidth=!, labelindent=\parindent]
    \item no point is allowed left of the line,
    \item the area enclosed between the found line and the line marking the minimum x-value, $x'_{\rm min}$, and maximum y-value, $y'_{\rm max}$, (shown as dashed lines in Fig. \ref{fig_PointCloudEx}) shall be maximised. Henceforth, we will call this area the `empty triangle'.
\end{enumerate}
From the second condition we can conclude, that the line must go through at least one point of the point cloud.
Otherwise we can always find a line parallel to the chosen one, that has all points to its right and produces a larger empty triangle.
Therefore, we start by determining which points out of our point cloud are relevant for our calculations.
These are all the points, through which we can put a line such that condition (1) is fulfilled.
They are marked with red `x's in Fig. \ref{fig_PointCloudEx}.

A linear function through a Point $P = (x'_{\rm P}, y'_{\rm P})$ can be written as:
\begin{align}
    y'_{\rm P} = a + b x'_{\rm P}.
\end{align}
Solving this equation for $a$ we can now express the area, $A$, of the empty triangle depending on the slope $b$:
\begin{align}
    A = \frac{1}{2}\left(y'_{\rm max} - y'_{\rm P}-b(x'_{\rm min} - x'_{\rm P})\right)\left( \frac{y'_{\rm max}-y'_{\rm P}}{b} + x'_{\rm P} - x'_{\rm min} \right).
\end{align}
It is easy to compute that for $A > 0$ this function has one global minimum at $b_{\rm min} = \frac{y'_{\rm max}-y'_{\rm P}}{x'_{\rm P}- x'_{\rm min}}$.
It is monotonic decreasing for $0 < b < b_{\rm min}$ and monotonic increasing for $b_{\rm min} < b$.

Due to condition (1) the minimum possible $b$ for a line through one of the relevant points is the one for which the line goes through its right neighbour and the maximum possible $b$ is the one for which its line goes through its left neighbour.
For the relevant points with the lowest and highest $x'$ only the line to one neighbour (the only one they have) has to be taken into account since all other possible lines with positive $b$ that fulfil (1) lead to a smaller empty triangle.
Therefore, by comparing the areas of the empty triangles created by the connections between all the neighbouring relevant points the function fulfilling both (1) and (2) can be found.
The connection producing the largest area is the solution.

Analogously a minimum function can be found by maximizing the triangle between $y' = y'_{\rm min}$ and $x' = x'_{\rm max}$ and the target function.
For a negative slope, $b$, $x' = x'_{\rm max}$ and $y' = y'_{\rm max}$ can be used to find the maximum function and $x' = x'_{\rm min}$ and $y' = y'_{\rm min}$ can be used to find the minimum function.

\bsp	
\label{lastpage}
\end{document}

%% file: tables/LatexTable1.tex
\begin{tabular}{lrrrrrrrrrrrrrrr}\hline
Name & \multicolumn{1}{c}{$M(t)$} & \multicolumn{1}{c}{$R_{\rm p}$} & \multicolumn{1}{c}{$R_{\rm ap}$} & \multicolumn{1}{c}{$\alpha_3$} & \multicolumn{1}{c}{$M_{\rm ini}$} & {$[Fe/H]$} & \multicolumn{1}{c}{$\sigma_{[Fe/H]}$} & \multicolumn{1}{c}{$N_{\rm SN}$} & \multicolumn{1}{c}{$n_{\rm SN}$} & \multicolumn{1}{c}{$N_{\rm SN}^{\rm exp}$} & \multicolumn{1}{c}{$t_{\rm end}^{\rm SF}$} \\
& \multicolumn{1}{c}{$[10^5 M_\odot]$} & \multicolumn{1}{c}{$[\rm kpc]$} & \multicolumn{1}{c}{$[\rm kpc]$} && \multicolumn{1}{c}{$[10^5 M_\odot]$} &&&& \multicolumn{1}{c}{$[M_\odot^{-1}]$} && \multicolumn{1}{c}{$[{\rm Myr}]$} \\
\hline
47 Tuc&     8.07&     5.47&     7.44&     1.75&    27.50&   -0.747&    0.033&{$3.25\times 10^{3}$}&{$1.18\times 10^{-3}$}&{$6.41\times 10^{4}$}&      3.5\\
NGC 288&     0.98&     2.01&    12.26&     1.83&    17.79&   -1.226&    0.037&{$7.82\times 10^{2}$}&{$4.39\times 10^{-4}$}&{$4.06\times 10^{4}$}&      3.4\\
NGC 362&     3.36&     1.02&    12.41&     1.65&    93.75&   -1.213&    0.074&{$8.52\times 10^{3}$}&{$9.09\times 10^{-4}$}&{$2.30\times 10^{5}$}&      3.5\\
NGC 1851&     2.81&     0.85&    19.20&     1.61&   136.44&   -1.157&    0.046&{$8.74\times 10^{3}$}&{$6.41\times 10^{-4}$}&{$3.38\times 10^{5}$}&      3.4\\
NGC 1904&     1.56&     0.82&    19.51&     1.62&   142.09&   -1.550&    0.027&{$2.16\times 10^{3}$}&{$1.52\times 10^{-4}$}&{$3.52\times 10^{5}$}&      3.5\\
NGC 2419&    14.30&    15.89&    92.20&     1.78&    42.97&   -2.095&    0.032&{$2.21\times 10^{2}$}&{$5.14\times 10^{-5}$}&{$1.01\times 10^{5}$}&      3.5\\
NGC 2808&     8.69&     0.97&    14.76&     1.61&   127.73&   -1.120&    0.035&{$6.77\times 10^{3}$}&{$5.30\times 10^{-4}$}&{$3.16\times 10^{5}$}&      3.4\\
NGC 3201&     1.41&     8.20&    24.15&     2.00&     4.32&   -1.496&    0.044&{$1.21\times 10^{2}$}&{$2.81\times 10^{-4}$}&{$8.75\times 10^{3}$}&      3.4\\
NGC 4590&     1.32&     8.95&    29.51&     2.06&     3.72&   -2.255&    0.053&{$2.19\times 10^{1}$}&{$5.89\times 10^{-5}$}&{$7.15\times 10^{3}$}&      3.4\\
NGC 4833&     1.80&     0.90&     7.68&     1.67&   118.88&   -2.070&    0.013&{$2.63\times 10^{2}$}&{$2.21\times 10^{-5}$}&{$2.91\times 10^{5}$}&      3.5\\
NGC 5024&     4.17&     9.11&    22.21&     1.92&    11.69&   -1.995&    0.071&{$1.68\times 10^{2}$}&{$1.44\times 10^{-4}$}&{$2.53\times 10^{4}$}&      3.4\\
NGC 5053&     0.73&    10.33&    17.74&     2.12&     2.20&   -2.450&    0.041&{$6.38\times 10^{0}$}&{$2.91\times 10^{-5}$}&{$3.93\times 10^{3}$}&      3.4\\
NGC 5139&    33.40&     1.33&     6.99&     1.62&   178.36&   -1.647&    0.271&{$2.32\times 10^{4}$}&{$1.30\times 10^{-3}$}&{$4.43\times 10^{5}$}&      3.5\\
NGC 5272&     3.60&     5.44&    15.16&     1.88&    12.17&   -1.391&    0.097&{$9.66\times 10^{2}$}&{$7.93\times 10^{-4}$}&{$2.69\times 10^{4}$}&      3.5\\
NGC 5286&     3.59&     1.17&    13.32&     1.72&    68.41&   -1.727&    0.103&{$2.66\times 10^{3}$}&{$3.89\times 10^{-4}$}&{$1.65\times 10^{5}$}&      3.4\\
NGC 5466&     0.55&     6.80&    53.55&     2.10&     2.07&   -1.865&    0.075&{$4.26\times 10^{1}$}&{$2.05\times 10^{-4}$}&{$3.81\times 10^{3}$}&      3.4\\
NGC 5634&     2.20&     3.97&    23.76&     1.94&     9.09&   -1.869&    0.081&{$20\times 10^{2}$}&{$2.20\times 10^{-4}$}&{$1.94\times 10^{4}$}&      3.4\\
NGC 5694&     3.67&     3.90&    65.74&     1.91&    13.10&   -2.017&    0.046&{$1.16\times 10^{2}$}&{$8.84\times 10^{-5}$}&{$2.86\times 10^{4}$}&      3.4\\
NGC 5824&     8.49&    12.48&    33.72&     1.85&    23.77&   -2.174&    0.058&{$1.85\times 10^{2}$}&{$7.78\times 10^{-5}$}&{$5.38\times 10^{4}$}&      3.5\\
NGC 5904&     3.68&     2.89&    24.03&     1.83&    17.63&   -1.259&    0.041&{$7.96\times 10^{2}$}&{$4.51\times 10^{-4}$}&{$4.02\times 10^{4}$}&      3.4\\
NGC 5986&     3.31&     0.70&     5.08&     1.56&   253.81&   -1.527&    0.061&{$9.21\times 10^{3}$}&{$3.63\times 10^{-4}$}&{$6.34\times 10^{5}$}&      3.4\\
NGC 6093&     2.64&     0.36&     3.53&     1.38&  1467.87&   -1.789&    0.014&{$6.67\times 10^{3}$}&{$4.54\times 10^{-5}$}&{$3.67\times 10^{6}$}&      3.5\\
NGC 6121&     0.93&     0.62&     6.20&     1.51&   325.37&   -1.166&    0.050&{$2.22\times 10^{4}$}&{$6.82\times 10^{-4}$}&{$8.14\times 10^{5}$}&      3.4\\
NGC 6139&     3.48&     1.37&     3.58&     1.71&    64.92&   -1.593&    0.033&{$1.09\times 10^{3}$}&{$1.68\times 10^{-4}$}&{$1.57\times 10^{5}$}&      3.4\\
NGC 6171&     0.81&     1.16&     3.74&     1.66&    75.02&   -0.949&    0.047&{$7.93\times 10^{3}$}&{$1.06\times 10^{-3}$}&{$1.83\times 10^{5}$}&      3.5\\
NGC 6205&     4.69&     1.56&     8.32&     1.74&    47.26&   -1.443&    0.101&{$3.47\times 10^{3}$}&{$7.33\times 10^{-4}$}&{$1.13\times 10^{5}$}&      3.4\\
NGC 6218&     0.87&     2.37&     4.80&     1.84&    16.64&   -1.315&    0.029&{$4.66\times 10^{2}$}&{$2.80\times 10^{-4}$}&{$3.77\times 10^{4}$}&      3.4\\
NGC 6229&     2.88&     2.05&    30.94&     1.80&    22.35&   -1.129&    0.044&{$1.46\times 10^{3}$}&{$6.54\times 10^{-4}$}&{$5.17\times 10^{4}$}&      3.5\\
NGC 6254&     1.89&     1.98&     4.59&     1.80&    27.25&   -1.559&    0.049&{$7.37\times 10^{2}$}&{$2.71\times 10^{-4}$}&{$6.33\times 10^{4}$}&      3.4\\
NGC 6266&     6.90&     0.83&     2.36&     1.55&   224.31&   -1.075&    0.041&{$1.55\times 10^{4}$}&{$6.89\times 10^{-4}$}&{$5.59\times 10^{5}$}&      3.4\\
NGC 6273&     6.57&     1.22&     3.34&     1.68&    95.00&   -1.612&    0.161&{$7.63\times 10^{3}$}&{$8.03\times 10^{-4}$}&{$2.32\times 10^{5}$}&      3.4\\
NGC 6341&     3.12&     1.00&    10.57&     1.71&    92.11&   -2.239&    0.083&{$8.85\times 10^{2}$}&{$9.61\times 10^{-5}$}&{$2.23\times 10^{5}$}&      3.5\\
NGC 6362&     1.16&     2.54&     5.16&     1.84&    15.32&   -1.092&    0.017&{$4.21\times 10^{2}$}&{$2.75\times 10^{-4}$}&{$3.47\times 10^{4}$}&      3.4\\
NGC 6366&     0.43&     1.97&     5.31&     1.79&    17.45&   -0.555&    0.071&{$6.92\times 10^{3}$}&{$3.97\times 10^{-3}$}&{$3.93\times 10^{4}$}&      4.9\\
NGC 6388&    11.30&     1.11&     3.78&     1.59&   110.26&   -0.428&    0.054&{$4.45\times 10^{4}$}&{$4.03\times 10^{-3}$}&{$2.66\times 10^{5}$}&      4.2\\
NGC 6397&     0.90&     2.63&     6.23&     1.91&    12.37&   -1.994&    0.028&{$7.02\times 10^{1}$}&{$5.67\times 10^{-5}$}&{$2.69\times 10^{4}$}&      3.5\\
NGC 6402&     7.39&     0.63&     4.36&     1.50&   362.45&   -1.130&    0.053&{$2.85\times 10^{4}$}&{$7.86\times 10^{-4}$}&{$9.07\times 10^{5}$}&      3.4\\
NGC 6441&    12.50&     1.00&     3.91&     1.57&   126.76&   -0.334&    0.079&{$9.31\times 10^{4}$}&{$7.35\times 10^{-3}$}&{$3.04\times 10^{5}$}&      5.9\\
NGC 6535&     0.13&     0.97&     4.50&     1.68&   103.20&   -1.963&    0.035&{$7.86\times 10^{2}$}&{$7.61\times 10^{-5}$}&{$2.52\times 10^{5}$}&      3.5\\
NGC 6553&     4.45&     1.26&     2.32&     1.66&    49.39&   -0.151&    0.047&{$3.28\times 10^{4}$}&{$6.64\times 10^{-3}$}&{$1.08\times 10^{5}$}&      6.3\\
NGC 6569&     2.28&     1.88&     2.97&     1.73&    36.03&   -0.867&    0.055&{$5.39\times 10^{3}$}&{$1.49\times 10^{-3}$}&{$8.54\times 10^{4}$}&      3.6\\
NGC 6626&     2.84&     0.58&     2.89&     1.49&   452.45&   -1.287&    0.075&{$3.51\times 10^{4}$}&{$7.77\times 10^{-4}$}&{$1.13\times 10^{6}$}&      3.4\\
NGC 6656&     4.09&     2.95&     9.48&     1.86&    19.75&   -1.803&    0.132&{$8.31\times 10^{2}$}&{$4.21\times 10^{-4}$}&{$4.46\times 10^{4}$}&      3.4\\
NGC 6681&     1.20&     0.85&     4.93&     1.62&   149.57&   -1.633&    0.028&{$1.95\times 10^{3}$}&{$1.30\times 10^{-4}$}&{$3.70\times 10^{5}$}&      3.5\\
NGC 6715&    16.20&    12.50&    36.59&     1.74&    50.19&   -1.559&    0.183&{$5.21\times 10^{3}$}&{$1.04\times 10^{-3}$}&{$1.20\times 10^{5}$}&      3.5\\
NGC 6752&     2.32&     3.23&     5.37&     1.87&    14.72&   -1.583&    0.034&{$2.61\times 10^{2}$}&{$1.77\times 10^{-4}$}&{$3.29\times 10^{4}$}&      3.4\\
NGC 6809&     1.88&     1.59&     5.54&     1.79&    36.99&   -1.934&    0.045&{$3.87\times 10^{2}$}&{$1.05\times 10^{-4}$}&{$8.66\times 10^{4}$}&      3.5\\
NGC 6838&     0.63&     4.75&     7.08&     1.95&     4.46&   -0.736&    0.039&{$6.38\times 10^{2}$}&{$1.43\times 10^{-3}$}&{$9.11\times 10^{3}$}&      3.8\\
NGC 6864&     4.10&     1.80&    17.51&     1.76&    33.22&   -1.164&    0.059&{$2.69\times 10^{3}$}&{$8.10\times 10^{-4}$}&{$7.84\times 10^{4}$}&      3.5\\
NGC 7078&     4.99&     3.57&    10.40&     1.88&    19.26&   -2.287&    0.053&{$1.05\times 10^{2}$}&{$5.48\times 10^{-5}$}&{$4.28\times 10^{4}$}&      3.5\\
NGC 7089&     5.20&     0.56&    16.93&     1.50&   417.15&   -1.399&    0.021&{$6.98\times 10^{3}$}&{$1.67\times 10^{-4}$}&{$1.05\times 10^{6}$}&      3.5\\
NGC 7099&     1.39&     1.49&     8.17&     1.82&    35.18&   -2.356&    0.037&{$1.15\times 10^{2}$}&{$3.26\times 10^{-5}$}&{$8.12\times 10^{4}$}&      3.5\\
Terzan 1&     2.72&     0.24&     1.48&     1.22&  5089.44&   -1.263&    0.037&{$2.05\times 10^{5}$}&{$4.03\times 10^{-4}$}&{$1.23\times 10^{7}$}&      3.4\\
Terzan 5&     7.60&     0.89&     2.96&     1.56&   136.44&   -0.092&    0.295&{$7.01\times 10^{5}$}&{$5.14\times 10^{-2}$}&{$3.27\times 10^{5}$}& {-} \\
Terzan 8&     0.58&     0.24&     1.65&     1.29&  4604.79&   -2.255&    0.098&{$5.05\times 10^{4}$}&{$1.10\times 10^{-4}$}&{$1.13\times 10^{7}$}&      3.5\\
\hline
\end{tabular}

%% file: IronSpreadGC2.bbl
\begin{thebibliography}{}
\makeatletter
\relax
\def\mn@urlcharsother{\let\do\@makeother \do\$\do\&\do\#\do\^\do\_\do\%\do\~}
\def\mn@doi{\begingroup\mn@urlcharsother \@ifnextchar [ {\mn@doi@}
  {\mn@doi@[]}}
\def\mn@doi@[#1]#2{\def\@tempa{#1}\ifx\@tempa\@empty \href
  {http://dx.doi.org/#2} {doi:#2}\else \href {http://dx.doi.org/#2} {#1}\fi
  \endgroup}
\def\mn@eprint#1#2{\mn@eprint@#1:#2::\@nil}
\def\mn@eprint@arXiv#1{\href {http://arxiv.org/abs/#1} {{\tt arXiv:#1}}}
\def\mn@eprint@dblp#1{\href {http://dblp.uni-trier.de/rec/bibtex/#1.xml}
  {dblp:#1}}
\def\mn@eprint@#1:#2:#3:#4\@nil{\def\@tempa {#1}\def\@tempb {#2}\def\@tempc
  {#3}\ifx \@tempc \@empty \let \@tempc \@tempb \let \@tempb \@tempa \fi \ifx
  \@tempb \@empty \def\@tempb {arXiv}\fi \@ifundefined
  {mn@eprint@\@tempb}{\@tempb:\@tempc}{\expandafter \expandafter \csname
  mn@eprint@\@tempb\endcsname \expandafter{\@tempc}}}

\bibitem[\protect\citeauthoryear{{Aarseth}, {Tout}  \& {Mardling}}{{Aarseth}
  et~al.}{2008}]{2008LNP...760.....A}
{Aarseth} S.~J.,  {Tout} C.~A.,   {Mardling} R.~A.,  2008, {The Cambridge
  N-Body Lectures}.
 Vol. 760, \mn@doi{10.1007/978-1-4020-8431-7, }

\bibitem[\protect\citeauthoryear{{Andr{\'e}}, {Di Francesco}, {Ward-Thompson},
  {Inutsuka}, {Pudritz}  \& {Pineda}}{{Andr{\'e}}
  et~al.}{2014}]{2014prpl.conf...27A}
{Andr{\'e}} P.,  {Di Francesco} J.,  {Ward-Thompson} D.,  {Inutsuka} S.~I.,
  {Pudritz} R.~E.,   {Pineda} J.~E.,  2014, in {Beuther} H.,  {Klessen} R.~S.,
  {Dullemond} C.~P.,   {Henning} T.,  eds, Protostars and Planets VI. p.~27
  (\mn@eprint {arXiv} {1312.6232}),
  \mn@doi{10.2458/azu_uapress_9780816531240-ch002}

\bibitem[\protect\citeauthoryear{{Ashman} \& {Zepf}}{{Ashman} \&
  {Zepf}}{2001}]{2001AJ....122.1888A}
{Ashman} K.~M.,  {Zepf} S.~E.,  2001, \mn@doi [\aj] {10.1086/323133}, \href
  {https://ui.adsabs.harvard.edu/abs/2001AJ....122.1888A} {122, 1888}

\bibitem[\protect\citeauthoryear{{Asplund}, {Grevesse}, {Sauval}  \&
  {Scott}}{{Asplund} et~al.}{2009}]{2009ARA&A..47..481A}
{Asplund} M.,  {Grevesse} N.,  {Sauval} A.~J.,   {Scott} P.,  2009, \mn@doi
  [\araa] {10.1146/annurev.astro.46.060407.145222}, \href
  {https://ui.adsabs.harvard.edu/abs/2009ARA&A..47..481A} {47, 481}

\bibitem[\protect\citeauthoryear{{Bailin}}{{Bailin}}{2019}]{2019ApJS..245....5B}
{Bailin} J.,  2019, \mn@doi [\apjs] {10.3847/1538-4365/ab4812}, \href
  {https://ui.adsabs.harvard.edu/abs/2019ApJS..245....5B} {245, 5}

\bibitem[\protect\citeauthoryear{{Bailin} \& {von Klar}}{{Bailin} \& {von
  Klar}}{2022}]{2022ApJ...925...36B}
{Bailin} J.,  {von Klar} R.,  2022, \mn@doi [\apj] {10.3847/1538-4357/ac347d},
  \href {https://ui.adsabs.harvard.edu/abs/2022ApJ...925...36B} {925, 36}

\bibitem[\protect\citeauthoryear{{Banerjee} \& {Kroupa}}{{Banerjee} \&
  {Kroupa}}{2011}]{2011ApJ...741L..12B}
{Banerjee} S.,  {Kroupa} P.,  2011, \mn@doi [\apjl]
  {10.1088/2041-8205/741/1/L12}, \href
  {https://ui.adsabs.harvard.edu/abs/2011ApJ...741L..12B} {741, L12}

\bibitem[\protect\citeauthoryear{{Banerjee} \& {Kroupa}}{{Banerjee} \&
  {Kroupa}}{2018}]{2018ASSL..424..143B}
{Banerjee} S.,  {Kroupa} P.,  2018, {Formation of Very Young Massive Clusters
  and Implications for Globular Clusters}.
p.~143, \mn@doi{10.1007/978-3-319-22801-3_6}

\bibitem[\protect\citeauthoryear{{Basinger}, {Kochanek}, {Adams}, {Dai}  \&
  {Stanek}}{{Basinger} et~al.}{2020}]{2020arXiv200715658B}
{Basinger} C.~M.,  {Kochanek} C.~S.,  {Adams} S.~M.,  {Dai} X.,   {Stanek}
  K.~Z.,  2020, arXiv e-prints, \href
  {https://ui.adsabs.harvard.edu/abs/2020arXiv200715658B} {p. arXiv:2007.15658}

\bibitem[\protect\citeauthoryear{{Bastian} \& {Lardo}}{{Bastian} \&
  {Lardo}}{2015}]{2015MNRAS.453..357B}
{Bastian} N.,  {Lardo} C.,  2015, \mn@doi [\mnras] {10.1093/mnras/stv1661},
  \href {https://ui.adsabs.harvard.edu/abs/2015MNRAS.453..357B} {453, 357}

\bibitem[\protect\citeauthoryear{{Bastian} \& {Strader}}{{Bastian} \&
  {Strader}}{2014}]{2014MNRAS.443.3594B}
{Bastian} N.,  {Strader} J.,  2014, \mn@doi [\mnras] {10.1093/mnras/stu1407},
  \href {https://ui.adsabs.harvard.edu/abs/2014MNRAS.443.3594B} {443, 3594}

\bibitem[\protect\citeauthoryear{{Baumgardt}}{{Baumgardt}}{2001}]{2001MNRAS.325.1323B}
{Baumgardt} H.,  2001, \mn@doi [\mnras] {10.1046/j.1365-8711.2001.04272.x},
  \href {https://ui.adsabs.harvard.edu/abs/2001MNRAS.325.1323B} {325, 1323}

\bibitem[\protect\citeauthoryear{{Baumgardt} \& {Makino}}{{Baumgardt} \&
  {Makino}}{2003}]{2003MNRAS.340..227B}
{Baumgardt} H.,  {Makino} J.,  2003, \mn@doi [\mnras]
  {10.1046/j.1365-8711.2003.06286.x}, \href
  {https://ui.adsabs.harvard.edu/abs/2003MNRAS.340..227B} {340, 227}

\bibitem[\protect\citeauthoryear{{Baumgardt}, {Kroupa}  \&
  {Parmentier}}{{Baumgardt} et~al.}{2008}]{2008MNRAS.384.1231B}
{Baumgardt} H.,  {Kroupa} P.,   {Parmentier} G.,  2008, \mn@doi [\mnras]
  {10.1111/j.1365-2966.2007.12811.x}, \href
  {https://ui.adsabs.harvard.edu/abs/2008MNRAS.384.1231B} {384, 1231}

\bibitem[\protect\citeauthoryear{{Baumgardt}, {Hilker}, {Sollima}  \&
  {Bellini}}{{Baumgardt} et~al.}{2019}]{2019MNRAS.482.5138B}
{Baumgardt} H.,  {Hilker} M.,  {Sollima} A.,   {Bellini} A.,  2019, \mn@doi
  [\mnras] {10.1093/mnras/sty2997}, \href
  {https://ui.adsabs.harvard.edu/abs/2019MNRAS.482.5138B} {482, 5138}

\bibitem[\protect\citeauthoryear{{Bekki} \& {Freeman}}{{Bekki} \&
  {Freeman}}{2003}]{2003MNRAS.346L..11B}
{Bekki} K.,  {Freeman} K.~C.,  2003, \mn@doi [\mnras]
  {10.1046/j.1365-2966.2003.07275.x}, \href
  {https://ui.adsabs.harvard.edu/abs/2003MNRAS.346L..11B} {346, L11}

\bibitem[\protect\citeauthoryear{{Bekki}, {Je{\v{r}}{\'a}bkov{\'a}}  \&
  {Kroupa}}{{Bekki} et~al.}{2017}]{2017MNRAS.471.2242B}
{Bekki} K.,  {Je{\v{r}}{\'a}bkov{\'a}} T.,   {Kroupa} P.,  2017, \mn@doi
  [\mnras] {10.1093/mnras/stx1609}, \href
  {https://ui.adsabs.harvard.edu/abs/2017MNRAS.471.2242B} {471, 2242}

\bibitem[\protect\citeauthoryear{{Brinkmann}, {Banerjee}, {Motwani}  \&
  {Kroupa}}{{Brinkmann} et~al.}{2017}]{2017A&A...600A..49B}
{Brinkmann} N.,  {Banerjee} S.,  {Motwani} B.,   {Kroupa} P.,  2017, \mn@doi
  [\aap] {10.1051/0004-6361/201629312}, \href
  {https://ui.adsabs.harvard.edu/abs/2017A&A...600A..49B} {600, A49}

\bibitem[\protect\citeauthoryear{{Calura}, {Few}, {Romano}  \&
  {D'Ercole}}{{Calura} et~al.}{2015}]{2015ApJ...814L..14C}
{Calura} F.,  {Few} C.~G.,  {Romano} D.,   {D'Ercole} A.,  2015, \mn@doi
  [\apjl] {10.1088/2041-8205/814/1/L14}, \href
  {https://ui.adsabs.harvard.edu/abs/2015ApJ...814L..14C} {814, L14}

\bibitem[\protect\citeauthoryear{{Carney}}{{Carney}}{1996}]{1996PASP..108..900C}
{Carney} B.~W.,  1996, \mn@doi [\pasp] {10.1086/133811}, \href
  {https://ui.adsabs.harvard.edu/abs/1996PASP..108..900C} {108, 900}

\bibitem[\protect\citeauthoryear{{Carretta} et~al.,}{{Carretta}
  et~al.}{2009a}]{2009A&A...505..117C}
{Carretta} E.,  et~al., 2009a, \mn@doi [\aap] {10.1051/0004-6361/200912096},
  \href {https://ui.adsabs.harvard.edu/abs/2009A&A...505..117C} {505, 117}

\bibitem[\protect\citeauthoryear{{Carretta}, {Bragaglia}, {Gratton}, {D'Orazi}
  \& {Lucatello}}{{Carretta} et~al.}{2009b}]{2009A&A...508..695C}
{Carretta} E.,  {Bragaglia} A.,  {Gratton} R.,  {D'Orazi} V.,   {Lucatello} S.,
   2009b, \mn@doi [\aap] {10.1051/0004-6361/200913003}, \href
  {https://ui.adsabs.harvard.edu/abs/2009A&A...508..695C} {508, 695}

\bibitem[\protect\citeauthoryear{{Carretta}, {Lucatello}, {Gratton},
  {Bragaglia}  \& {D'Orazi}}{{Carretta} et~al.}{2011}]{2011A&A...533A..69C}
{Carretta} E.,  {Lucatello} S.,  {Gratton} R.~G.,  {Bragaglia} A.,   {D'Orazi}
  V.,  2011, \mn@doi [\aap] {10.1051/0004-6361/201117269}, \href
  {https://ui.adsabs.harvard.edu/abs/2011A&A...533A..69C} {533, A69}

\bibitem[\protect\citeauthoryear{{Carretta}, {Bragaglia}, {Gratton},
  {Lucatello}  \& {D'Orazi}}{{Carretta} et~al.}{2012}]{2012ApJ...750L..14C}
{Carretta} E.,  {Bragaglia} A.,  {Gratton} R.~G.,  {Lucatello} S.,   {D'Orazi}
  V.,  2012, \mn@doi [\apjl] {10.1088/2041-8205/750/1/L14}, \href
  {https://ui.adsabs.harvard.edu/abs/2012ApJ...750L..14C} {750, L14}

\bibitem[\protect\citeauthoryear{{Carretta} et~al.,}{{Carretta}
  et~al.}{2013}]{2013A&A...557A.138C}
{Carretta} E.,  et~al., 2013, \mn@doi [\aap] {10.1051/0004-6361/201321905},
  \href {https://ui.adsabs.harvard.edu/abs/2013A&A...557A.138C} {557, A138}

\bibitem[\protect\citeauthoryear{{Carretta} et~al.,}{{Carretta}
  et~al.}{2014}]{2014A&A...564A..60C}
{Carretta} E.,  et~al., 2014, \mn@doi [\aap] {10.1051/0004-6361/201323321},
  \href {https://ui.adsabs.harvard.edu/abs/2014A&A...564A..60C} {564, A60}

\bibitem[\protect\citeauthoryear{{Carretta} et~al.,}{{Carretta}
  et~al.}{2015}]{2015A&A...578A.116C}
{Carretta} E.,  et~al., 2015, \mn@doi [\aap] {10.1051/0004-6361/201525951},
  \href {https://ui.adsabs.harvard.edu/abs/2015A&A...578A.116C} {578, A116}

\bibitem[\protect\citeauthoryear{{Chon}, {Omukai}  \& {Schneider}}{{Chon}
  et~al.}{2021}]{2021arXiv210304997C}
{Chon} S.,  {Omukai} K.,   {Schneider} R.,  2021, arXiv e-prints, \href
  {https://ui.adsabs.harvard.edu/abs/2021arXiv210304997C} {p. arXiv:2103.04997}

\bibitem[\protect\citeauthoryear{{Cohen}, {Bellini}, {Casagrande}, {Brown},
  {Correnti}  \& {Kalirai}}{{Cohen} et~al.}{2021}]{2021arXiv210908708C}
{Cohen} R.~E.,  {Bellini} A.,  {Casagrande} L.,  {Brown} T.~M.,  {Correnti} M.,
    {Kalirai} J.~S.,  2021, arXiv e-prints, \href
  {https://ui.adsabs.harvard.edu/abs/2021arXiv210908708C} {p. arXiv:2109.08708}

\bibitem[\protect\citeauthoryear{{Corbelli}}{{Corbelli}}{2003}]{2003MNRAS.342..199C}
{Corbelli} E.,  2003, \mn@doi [\mnras] {10.1046/j.1365-8711.2003.06531.x},
  \href {https://ui.adsabs.harvard.edu/abs/2003MNRAS.342..199C} {342, 199}

\bibitem[\protect\citeauthoryear{{D'Antona}, {Vesperini}, {D'Ercole},
  {Ventura}, {Milone}, {Marino}  \& {Tailo}}{{D'Antona}
  et~al.}{2016}]{2016MNRAS.458.2122D}
{D'Antona} F.,  {Vesperini} E.,  {D'Ercole} A.,  {Ventura} P.,  {Milone} A.~P.,
   {Marino} A.~F.,   {Tailo} M.,  2016, \mn@doi [\mnras]
  {10.1093/mnras/stw387}, \href
  {https://ui.adsabs.harvard.edu/abs/2016MNRAS.458.2122D} {458, 2122}

\bibitem[\protect\citeauthoryear{{Dabringhausen}, {Kroupa}  \&
  {Baumgardt}}{{Dabringhausen} et~al.}{2009}]{2009MNRAS.394.1529D}
{Dabringhausen} J.,  {Kroupa} P.,   {Baumgardt} H.,  2009, \mn@doi [\mnras]
  {10.1111/j.1365-2966.2009.14425.x}, \href
  {https://ui.adsabs.harvard.edu/abs/2009MNRAS.394.1529D} {394, 1529}

\bibitem[\protect\citeauthoryear{{Dabringhausen}, {Kroupa}, {Pflamm-Altenburg}
  \& {Mieske}}{{Dabringhausen} et~al.}{2012}]{2012ApJ...747...72D}
{Dabringhausen} J.,  {Kroupa} P.,  {Pflamm-Altenburg} J.,   {Mieske} S.,  2012,
  \mn@doi [\apj] {10.1088/0004-637X/747/1/72}, \href
  {https://ui.adsabs.harvard.edu/abs/2012ApJ...747...72D} {747, 72}

\bibitem[\protect\citeauthoryear{{Dale}, {Ercolano}  \& {Bonnell}}{{Dale}
  et~al.}{2012}]{2012MNRAS.424..377D}
{Dale} J.~E.,  {Ercolano} B.,   {Bonnell} I.~A.,  2012, \mn@doi [\mnras]
  {10.1111/j.1365-2966.2012.21205.x}, \href
  {https://ui.adsabs.harvard.edu/abs/2012MNRAS.424..377D} {424, 377}

\bibitem[\protect\citeauthoryear{{Deason}, {Belokurov}  \& {Sanders}}{{Deason}
  et~al.}{2019}]{2019MNRAS.490.3426D}
{Deason} A.~J.,  {Belokurov} V.,   {Sanders} J.~L.,  2019, \mn@doi [\mnras]
  {10.1093/mnras/stz2793}, \href
  {https://ui.adsabs.harvard.edu/abs/2019MNRAS.490.3426D} {490, 3426}

\bibitem[\protect\citeauthoryear{{Dotter} et~al.,}{{Dotter}
  et~al.}{2010}]{2010ApJ...708..698D}
{Dotter} A.,  et~al., 2010, \mn@doi [\apj] {10.1088/0004-637X/708/1/698}, \href
  {https://ui.adsabs.harvard.edu/abs/2010ApJ...708..698D} {708, 698}

\bibitem[\protect\citeauthoryear{{Fall} \& {Zhang}}{{Fall} \&
  {Zhang}}{2001}]{2001ApJ...561..751F}
{Fall} S.~M.,  {Zhang} Q.,  2001, \mn@doi [\apj] {10.1086/323358}, \href
  {https://ui.adsabs.harvard.edu/abs/2001ApJ...561..751F} {561, 751}

\bibitem[\protect\citeauthoryear{{Farrell}, {Groh}, {Meynet}  \&
  {Eldridge}}{{Farrell} et~al.}{2022}]{2022MNRAS.tmp..580F}
{Farrell} E.,  {Groh} J.~H.,  {Meynet} G.,   {Eldridge} J.~J.,  2022, \mn@doi
  [\mnras] {10.1093/mnras/stac538}, \href
  {https://ui.adsabs.harvard.edu/abs/2022MNRAS.tmp..580F} {}

\bibitem[\protect\citeauthoryear{{Ferraro} et~al.,}{{Ferraro}
  et~al.}{2009}]{2009Natur.462..483F}
{Ferraro} F.~R.,  et~al., 2009, \mn@doi [\nat] {10.1038/nature08581}, \href
  {https://ui.adsabs.harvard.edu/abs/2009Natur.462..483F} {462, 483}

\bibitem[\protect\citeauthoryear{{Forbes}, {Spitler}, {Strader}, {Romanowsky},
  {Brodie}  \& {Foster}}{{Forbes} et~al.}{2011}]{2011MNRAS.413.2943F}
{Forbes} D.~A.,  {Spitler} L.~R.,  {Strader} J.,  {Romanowsky} A.~J.,  {Brodie}
  J.~P.,   {Foster} C.,  2011, \mn@doi [\mnras]
  {10.1111/j.1365-2966.2011.18373.x}, \href
  {https://ui.adsabs.harvard.edu/abs/2011MNRAS.413.2943F} {413, 2943}

\bibitem[\protect\citeauthoryear{{Gilmore} \& {Wyse}}{{Gilmore} \&
  {Wyse}}{1998}]{1998AJ....116..748G}
{Gilmore} G.,  {Wyse} R. F.~G.,  1998, \mn@doi [\aj] {10.1086/300459}, \href
  {https://ui.adsabs.harvard.edu/abs/1998AJ....116..748G} {116, 748}

\bibitem[\protect\citeauthoryear{{G{\"u}rkan}, {Freitag}  \&
  {Rasio}}{{G{\"u}rkan} et~al.}{2004}]{2004ApJ...604..632G}
{G{\"u}rkan} M.~A.,  {Freitag} M.,   {Rasio} F.~A.,  2004, \mn@doi [\apj]
  {10.1086/381968}, \href
  {https://ui.adsabs.harvard.edu/abs/2004ApJ...604..632G} {604, 632}

\bibitem[\protect\citeauthoryear{{Haghi}, {Safaei}, {Zonoozi}  \&
  {Kroupa}}{{Haghi} et~al.}{2020}]{2020ApJ...904...43H}
{Haghi} H.,  {Safaei} G.,  {Zonoozi} A.~H.,   {Kroupa} P.,  2020, \mn@doi
  [\apj] {10.3847/1538-4357/abbfb0}, \href
  {https://ui.adsabs.harvard.edu/abs/2020ApJ...904...43H} {904, 43}

\bibitem[\protect\citeauthoryear{{Han}, {Kimm}, {Katz}, {Devriendt}  \&
  {Slyz}}{{Han} et~al.}{2022}]{2022arXiv220705745H}
{Han} D.,  {Kimm} T.,  {Katz} H.,  {Devriendt} J.,   {Slyz} A.,  2022, arXiv
  e-prints, \href {https://ui.adsabs.harvard.edu/abs/2022arXiv220705745H} {p.
  arXiv:2207.05745}

\bibitem[\protect\citeauthoryear{{Harris}, {Harris}  \& {Alessi}}{{Harris}
  et~al.}{2013}]{2013ApJ...772...82H}
{Harris} W.~E.,  {Harris} G. L.~H.,   {Alessi} M.,  2013, \mn@doi [\apj]
  {10.1088/0004-637X/772/2/82}, \href
  {https://ui.adsabs.harvard.edu/abs/2013ApJ...772...82H} {772, 82}

\bibitem[\protect\citeauthoryear{{Heger}, {Woosley}, {Fryer}  \&
  {Langer}}{{Heger} et~al.}{2003a}]{2003fthp.conf....3H}
{Heger} A.,  {Woosley} S.~E.,  {Fryer} C.~L.,   {Langer} N.,  2003a, in
  {Hillebrandt} W.,  {Leibundgut} B.,  eds, From Twilight to Highlight: The
  Physics of Supernovae. p.~3 (\mn@eprint {arXiv} {astro-ph/0211062}),
  \mn@doi{10.1007/10828549_1}

\bibitem[\protect\citeauthoryear{{Heger}, {Fryer}, {Woosley}, {Langer}  \&
  {Hartmann}}{{Heger} et~al.}{2003b}]{2003ApJ...591..288H}
{Heger} A.,  {Fryer} C.~L.,  {Woosley} S.~E.,  {Langer} N.,   {Hartmann} D.~H.,
   2003b, \mn@doi [\apj] {10.1086/375341}, \href
  {https://ui.adsabs.harvard.edu/abs/2003ApJ...591..288H} {591, 288}

\bibitem[\protect\citeauthoryear{{Heggie} \& {Hut}}{{Heggie} \&
  {Hut}}{2003}]{2003gmbp.book.....H}
{Heggie} D.,  {Hut} P.,  2003, {The Gravitational Million-Body Problem: A
  Multidisciplinary Approach to Star Cluster Dynamics}

\bibitem[\protect\citeauthoryear{Hilker, Baumgardt, Sollima  \& Bellini}{Hilker
  et~al.}{2019}]{hilker_baumgardt_sollima_bellini_2019}
Hilker M.,  Baumgardt H.,  Sollima A.,   Bellini A.,  2019, \mn@doi
  [Proceedings of the International Astronomical Union]
  {10.1017/S1743921319006823}, 14, 451–454

\bibitem[\protect\citeauthoryear{{Jerabkova}, {Boffin}, {Beccari}, {de Marchi},
  {de Bruijne}  \& {Prusti}}{{Jerabkova} et~al.}{2021}]{2021A&A...647A.137J}
{Jerabkova} T.,  {Boffin} H. M.~J.,  {Beccari} G.,  {de Marchi} G.,  {de
  Bruijne} J. H.~J.,   {Prusti} T.,  2021, \mn@doi [\aap]
  {10.1051/0004-6361/202039949}, \href
  {https://ui.adsabs.harvard.edu/abs/2021A&A...647A.137J} {647, A137}

\bibitem[\protect\citeauthoryear{{Je{\v{r}}{\'a}bkov{\'a}}, {Kroupa},
  {Dabringhausen}, {Hilker}  \& {Bekki}}{{Je{\v{r}}{\'a}bkov{\'a}}
  et~al.}{2017}]{2017A&A...608A..53J}
{Je{\v{r}}{\'a}bkov{\'a}} T.,  {Kroupa} P.,  {Dabringhausen} J.,  {Hilker} M.,
   {Bekki} K.,  2017, \mn@doi [\aap] {10.1051/0004-6361/201731240}, \href
  {https://ui.adsabs.harvard.edu/abs/2017A&A...608A..53J} {608, A53}

\bibitem[\protect\citeauthoryear{{Je{\v{r}}{\'a}bkov{\'a}}, {Hasani Zonoozi},
  {Kroupa}, {Beccari}, {Yan}, {Vazdekis}  \& {Zhang}}{{Je{\v{r}}{\'a}bkov{\'a}}
  et~al.}{2018}]{2018A&A...620A..39J}
{Je{\v{r}}{\'a}bkov{\'a}} T.,  {Hasani Zonoozi} A.,  {Kroupa} P.,  {Beccari}
  G.,  {Yan} Z.,  {Vazdekis} A.,   {Zhang} Z.~Y.,  2018, \mn@doi [\aap]
  {10.1051/0004-6361/201833055}, \href
  {https://ui.adsabs.harvard.edu/abs/2018A&A...620A..39J} {620, A39}

\bibitem[\protect\citeauthoryear{{Joshi}, {Nave}  \& {Rasio}}{{Joshi}
  et~al.}{2001}]{2001ApJ...550..691J}
{Joshi} K.~J.,  {Nave} C.~P.,   {Rasio} F.~A.,  2001, \mn@doi [\apj]
  {10.1086/319771}, \href
  {https://ui.adsabs.harvard.edu/abs/2001ApJ...550..691J} {550, 691}

\bibitem[\protect\citeauthoryear{{Kacharov}, {Koch}  \& {McWilliam}}{{Kacharov}
  et~al.}{2013}]{2013A&A...554A..81K}
{Kacharov} N.,  {Koch} A.,   {McWilliam} A.,  2013, \mn@doi [\aap]
  {10.1051/0004-6361/201321392}, \href
  {https://ui.adsabs.harvard.edu/abs/2013A&A...554A..81K} {554, A81}

\bibitem[\protect\citeauthoryear{{Kafle}, {Sharma}, {Lewis}  \&
  {Bland-Hawthorn}}{{Kafle} et~al.}{2012}]{2012ApJ...761...98K}
{Kafle} P.~R.,  {Sharma} S.,  {Lewis} G.~F.,   {Bland-Hawthorn} J.,  2012,
  \mn@doi [\apj] {10.1088/0004-637X/761/2/98}, \href
  {https://ui.adsabs.harvard.edu/abs/2012ApJ...761...98K} {761, 98}

\bibitem[\protect\citeauthoryear{{Kalari}, {Carraro}, {Evans}  \&
  {Rubio}}{{Kalari} et~al.}{2018}]{2018ApJ...857..132K}
{Kalari} V.~M.,  {Carraro} G.,  {Evans} C.~J.,   {Rubio} M.,  2018, \mn@doi
  [\apj] {10.3847/1538-4357/aab609}, \href
  {https://ui.adsabs.harvard.edu/abs/2018ApJ...857..132K} {857, 132}

\bibitem[\protect\citeauthoryear{{Krause}, {Charbonnel}, {Bastian}  \&
  {Diehl}}{{Krause} et~al.}{2016}]{2016A&A...587A..53K}
{Krause} M. G.~H.,  {Charbonnel} C.,  {Bastian} N.,   {Diehl} R.,  2016,
  \mn@doi [\aap] {10.1051/0004-6361/201526685}, \href
  {https://ui.adsabs.harvard.edu/abs/2016A&A...587A..53K} {587, A53}

\bibitem[\protect\citeauthoryear{{Kravtsov}, {Dib}, {Calder{\'o}n}  \&
  {Belinch{\'o}n}}{{Kravtsov} et~al.}{2022}]{2022MNRAS.512.2936K}
{Kravtsov} V.,  {Dib} S.,  {Calder{\'o}n} F.~A.,   {Belinch{\'o}n} J.~A.,
  2022, \mn@doi [\mnras] {10.1093/mnras/stac716}, \href
  {https://ui.adsabs.harvard.edu/abs/2022MNRAS.512.2936K} {512, 2936}

\bibitem[\protect\citeauthoryear{{Kroupa}}{{Kroupa}}{2001}]{2001MNRAS.322..231K}
{Kroupa} P.,  2001, \mn@doi [\mnras] {10.1046/j.1365-8711.2001.04022.x}, \href
  {https://ui.adsabs.harvard.edu/abs/2001MNRAS.322..231K} {322, 231}

\bibitem[\protect\citeauthoryear{{Kroupa}}{{Kroupa}}{2002}]{2002Sci...295...82K}
{Kroupa} P.,  2002, \mn@doi [Science] {10.1126/science.1067524}, \href
  {https://ui.adsabs.harvard.edu/abs/2002Sci...295...82K} {295, 82}

\bibitem[\protect\citeauthoryear{{Kroupa}, {Tout}  \& {Gilmore}}{{Kroupa}
  et~al.}{1993}]{1993MNRAS.262..545K}
{Kroupa} P.,  {Tout} C.~A.,   {Gilmore} G.,  1993, \mn@doi [\mnras]
  {10.1093/mnras/262.3.545}, \href
  {https://ui.adsabs.harvard.edu/abs/1993MNRAS.262..545K} {262, 545}

\bibitem[\protect\citeauthoryear{{Kroupa}, {Weidner}, {Pflamm-Altenburg},
  {Thies}, {Dabringhausen}, {Marks}  \& {Maschberger}}{{Kroupa}
  et~al.}{2013}]{2013pss5.book..115K}
{Kroupa} P.,  {Weidner} C.,  {Pflamm-Altenburg} J.,  {Thies} I.,
  {Dabringhausen} J.,  {Marks} M.,   {Maschberger} T.,  2013, {The Stellar and
  Sub-Stellar Initial Mass Function of Simple and Composite Populations}.
Springer Netherlands, p.~115, \mn@doi{10.1007/978-94-007-5612-0_4}

\bibitem[\protect\citeauthoryear{{Kroupa}, {Subr}, {Jerabkova}  \&
  {Wang}}{{Kroupa} et~al.}{2020}]{2020MNRAS.498.5652K}
{Kroupa} P.,  {Subr} L.,  {Jerabkova} T.,   {Wang} L.,  2020, \mn@doi [\mnras]
  {10.1093/mnras/staa2276}, \href
  {https://ui.adsabs.harvard.edu/abs/2020MNRAS.498.5652K} {498, 5652}

\bibitem[\protect\citeauthoryear{{Lacchin}, {Calura}  \& {Vesperini}}{{Lacchin}
  et~al.}{2021}]{2021MNRAS.506.5951L}
{Lacchin} E.,  {Calura} F.,   {Vesperini} E.,  2021, \mn@doi [\mnras]
  {10.1093/mnras/stab2061}, \href
  {https://ui.adsabs.harvard.edu/abs/2021MNRAS.506.5951L} {506, 5951}

\bibitem[\protect\citeauthoryear{{Lada} \& {Lada}}{{Lada} \&
  {Lada}}{2003}]{2003ARA&A..41...57L}
{Lada} C.~J.,  {Lada} E.~A.,  2003, \mn@doi [\araa]
  {10.1146/annurev.astro.41.011802.094844}, \href
  {https://ui.adsabs.harvard.edu/abs/2003ARA&A..41...57L} {41, 57}

\bibitem[\protect\citeauthoryear{{Lamers}, {Baumgardt}  \& {Gieles}}{{Lamers}
  et~al.}{2010}]{2010MNRAS.409..305L}
{Lamers} H. J.~G.~L.~M.,  {Baumgardt} H.,   {Gieles} M.,  2010, \mn@doi
  [\mnras] {10.1111/j.1365-2966.2010.17309.x}, \href
  {https://ui.adsabs.harvard.edu/abs/2010MNRAS.409..305L} {409, 305}

\bibitem[\protect\citeauthoryear{{Lardo}, {Mucciarelli}  \& {Bastian}}{{Lardo}
  et~al.}{2016}]{2016MNRAS.457...51L}
{Lardo} C.,  {Mucciarelli} A.,   {Bastian} N.,  2016, \mn@doi [\mnras]
  {10.1093/mnras/stv2802}, \href
  {https://ui.adsabs.harvard.edu/abs/2016MNRAS.457...51L} {457, 51}

\bibitem[\protect\citeauthoryear{{Lardo}, {Salaris}, {Cassisi}  \&
  {Bastian}}{{Lardo} et~al.}{2022}]{2022arXiv220503323L}
{Lardo} C.,  {Salaris} M.,  {Cassisi} S.,   {Bastian} N.,  2022, arXiv
  e-prints, \href {https://ui.adsabs.harvard.edu/abs/2022arXiv220503323L} {p.
  arXiv:2205.03323}

\bibitem[\protect\citeauthoryear{{Mahani}, {Zonoozi}, {Haghi},
  {Je{\v{r}}{\'a}bkov{\'a}}, {Kroupa}  \& {Mieske}}{{Mahani}
  et~al.}{2021}]{2021MNRAS.502.5185M}
{Mahani} H.,  {Zonoozi} A.~H.,  {Haghi} H.,  {Je{\v{r}}{\'a}bkov{\'a}} T.,
  {Kroupa} P.,   {Mieske} S.,  2021, \mn@doi [\mnras] {10.1093/mnras/stab330},
  \href {https://ui.adsabs.harvard.edu/abs/2021MNRAS.502.5185M} {502, 5185}

\bibitem[\protect\citeauthoryear{{Maoz} \& {Graur}}{{Maoz} \&
  {Graur}}{2017}]{2017ApJ...848...25M}
{Maoz} D.,  {Graur} O.,  2017, \mn@doi [\apj] {10.3847/1538-4357/aa8b6e}, \href
  {https://ui.adsabs.harvard.edu/abs/2017ApJ...848...25M} {848, 25}

\bibitem[\protect\citeauthoryear{{Marigo}}{{Marigo}}{2001}]{2001A&A...370..194M}
{Marigo} P.,  2001, \mn@doi [\aap] {10.1051/0004-6361:20000247}, \href
  {https://ui.adsabs.harvard.edu/abs/2001A&A...370..194M} {370, 194}

\bibitem[\protect\citeauthoryear{{Marino} et~al.,}{{Marino}
  et~al.}{2015}]{2015MNRAS.450..815M}
{Marino} A.~F.,  et~al., 2015, \mn@doi [\mnras] {10.1093/mnras/stv420}, \href
  {https://ui.adsabs.harvard.edu/abs/2015MNRAS.450..815M} {450, 815}

\bibitem[\protect\citeauthoryear{{Marino} et~al.,}{{Marino}
  et~al.}{2018}]{2018ApJ...859...81M}
{Marino} A.~F.,  et~al., 2018, \mn@doi [\apj] {10.3847/1538-4357/aabdea}, \href
  {https://ui.adsabs.harvard.edu/abs/2018ApJ...859...81M} {859, 81}

\bibitem[\protect\citeauthoryear{{Marino} et~al.,}{{Marino}
  et~al.}{2019}]{2019MNRAS.487.3815M}
{Marino} A.~F.,  et~al., 2019, \mn@doi [\mnras] {10.1093/mnras/stz1415}, \href
  {https://ui.adsabs.harvard.edu/abs/2019MNRAS.487.3815M} {487, 3815}

\bibitem[\protect\citeauthoryear{{Marks} \& {Kroupa}}{{Marks} \&
  {Kroupa}}{2012}]{2012A&A...543A...8M}
{Marks} M.,  {Kroupa} P.,  2012, \mn@doi [\aap] {10.1051/0004-6361/201118231},
  \href {https://ui.adsabs.harvard.edu/abs/2012A&A...543A...8M} {543, A8}

\bibitem[\protect\citeauthoryear{{Marks}, {Kroupa}, {Dabringhausen}  \&
  {Pawlowski}}{{Marks} et~al.}{2012}]{2012MNRAS.422.2246M}
{Marks} M.,  {Kroupa} P.,  {Dabringhausen} J.,   {Pawlowski} M.~S.,  2012,
  \mn@doi [\mnras] {10.1111/j.1365-2966.2012.20767.x}, \href
  {https://ui.adsabs.harvard.edu/abs/2012MNRAS.422.2246M} {422, 2246}

\bibitem[\protect\citeauthoryear{{Marks}, {Kroupa}  \& {Dabringhausen}}{{Marks}
  et~al.}{2022}]{2022A&A...659A..96M}
{Marks} M.,  {Kroupa} P.,   {Dabringhausen} J.,  2022, \mn@doi [\aap]
  {10.1051/0004-6361/202141846}, \href
  {https://ui.adsabs.harvard.edu/abs/2022A&A...659A..96M} {659, A96}

\bibitem[\protect\citeauthoryear{Marquardt}{Marquardt}{1963}]{doi:10.1137/0111030}
Marquardt D.~W.,  1963, \mn@doi [Journal of the Society for Industrial and
  Applied Mathematics] {10.1137/0111030}, 11, 431

\bibitem[\protect\citeauthoryear{{Massari} et~al.,}{{Massari}
  et~al.}{2014}]{2014ApJ...795...22M}
{Massari} D.,  et~al., 2014, \mn@doi [\apj] {10.1088/0004-637X/795/1/22}, \href
  {https://ui.adsabs.harvard.edu/abs/2014ApJ...795...22M} {795, 22}

\bibitem[\protect\citeauthoryear{{McMillan}}{{McMillan}}{2011}]{2011MNRAS.414.2446M}
{McMillan} P.~J.,  2011, \mn@doi [\mnras] {10.1111/j.1365-2966.2011.18564.x},
  \href {https://ui.adsabs.harvard.edu/abs/2011MNRAS.414.2446M} {414, 2446}

\bibitem[\protect\citeauthoryear{{McMillan}}{{McMillan}}{2017}]{2017MNRAS.465...76M}
{McMillan} P.~J.,  2017, \mn@doi [\mnras] {10.1093/mnras/stw2759}, \href
  {https://ui.adsabs.harvard.edu/abs/2017MNRAS.465...76M} {465, 76}

\bibitem[\protect\citeauthoryear{{Megeath} et~al.,}{{Megeath}
  et~al.}{2016}]{2016AJ....151....5M}
{Megeath} S.~T.,  et~al., 2016, \mn@doi [\aj] {10.3847/0004-6256/151/1/5},
  \href {https://ui.adsabs.harvard.edu/abs/2016AJ....151....5M} {151, 5}

\bibitem[\protect\citeauthoryear{{Montecinos}, {Villanova}, {Mu{\v{n}}oz}  \&
  {Cort{\'e}s}}{{Montecinos} et~al.}{2021}]{2021arXiv210307014M}
{Montecinos} C.,  {Villanova} S.,  {Mu{\v{n}}oz} C.,   {Cort{\'e}s} C.~C.,
  2021, arXiv e-prints, \href
  {https://ui.adsabs.harvard.edu/abs/2021arXiv210307014M} {p. arXiv:2103.07014}

\bibitem[\protect\citeauthoryear{{Mucciarelli}, {Lapenna}, {Massari},
  {Pancino}, {Stetson}, {Ferraro}, {Lanzoni}  \& {Lardo}}{{Mucciarelli}
  et~al.}{2015}]{2015ApJ...809..128M}
{Mucciarelli} A.,  {Lapenna} E.,  {Massari} D.,  {Pancino} E.,  {Stetson}
  P.~B.,  {Ferraro} F.~R.,  {Lanzoni} B.,   {Lardo} C.,  2015, \mn@doi [\apj]
  {10.1088/0004-637X/809/2/128}, \href
  {https://ui.adsabs.harvard.edu/abs/2015ApJ...809..128M} {809, 128}

\bibitem[\protect\citeauthoryear{{Neustadt}, {Kochanek}, {Stanek}, {Basinger},
  {Jayasinghe}, {Garling}, {Adams}  \& {Gerke}}{{Neustadt}
  et~al.}{2021}]{2021arXiv210403318N}
{Neustadt} J.~M.~M.,  {Kochanek} C.~S.,  {Stanek} K.~Z.,  {Basinger} C.~M.,
  {Jayasinghe} T.,  {Garling} C.~T.,  {Adams} S.~M.,   {Gerke} J.,  2021, arXiv
  e-prints, \href {https://ui.adsabs.harvard.edu/abs/2021arXiv210403318N} {p.
  arXiv:2104.03318}

\bibitem[\protect\citeauthoryear{{Nomoto}, {Kobayashi}  \& {Tominaga}}{{Nomoto}
  et~al.}{2013}]{2013ARA&A..51..457N}
{Nomoto} K.,  {Kobayashi} C.,   {Tominaga} N.,  2013, \mn@doi [\araa]
  {10.1146/annurev-astro-082812-140956}, \href
  {https://ui.adsabs.harvard.edu/abs/2013ARA&A..51..457N} {51, 457}

\bibitem[\protect\citeauthoryear{{Pancino}, {Ferraro}, {Bellazzini}, {Piotto}
  \& {Zoccali}}{{Pancino} et~al.}{2000}]{2000ApJ...534L..83P}
{Pancino} E.,  {Ferraro} F.~R.,  {Bellazzini} M.,  {Piotto} G.,   {Zoccali} M.,
   2000, \mn@doi [\apjl] {10.1086/312658}, \href
  {https://ui.adsabs.harvard.edu/abs/2000ApJ...534L..83P} {534, L83}

\bibitem[\protect\citeauthoryear{{Parmentier}, {Jehin}, {Magain}, {Neuforge},
  {Noels}  \& {Thoul}}{{Parmentier} et~al.}{1999}]{1999A&A...352..138P}
{Parmentier} G.,  {Jehin} E.,  {Magain} P.,  {Neuforge} C.,  {Noels} A.,
  {Thoul} A.~A.,  1999, \aap, \href
  {https://ui.adsabs.harvard.edu/abs/1999A&A...352..138P} {352, 138}

\bibitem[\protect\citeauthoryear{{Pejcha} \& {Thompson}}{{Pejcha} \&
  {Thompson}}{2015}]{2015ApJ...801...90P}
{Pejcha} O.,  {Thompson} T.~A.,  2015, \mn@doi [\apj]
  {10.1088/0004-637X/801/2/90}, \href
  {https://ui.adsabs.harvard.edu/abs/2015ApJ...801...90P} {801, 90}

\bibitem[\protect\citeauthoryear{{Pflamm-Altenburg},
  {Gonz{\'a}lez-L{\'o}pezlira}  \& {Kroupa}}{{Pflamm-Altenburg}
  et~al.}{2013}]{2013MNRAS.435.2604P}
{Pflamm-Altenburg} J.,  {Gonz{\'a}lez-L{\'o}pezlira} R.~A.,   {Kroupa} P.,
  2013, \mn@doi [\mnras] {10.1093/mnras/stt1474}, \href
  {https://ui.adsabs.harvard.edu/abs/2013MNRAS.435.2604P} {435, 2604}

\bibitem[\protect\citeauthoryear{{Portegies Zwart} \& {Boekholt}}{{Portegies
  Zwart} \& {Boekholt}}{2018}]{2018CNSNS..61..160P}
{Portegies Zwart} S.~F.,  {Boekholt} T. C.~N.,  2018, \mn@doi [Communications
  in Nonlinear Science and Numerical Simulations]
  {10.1016/j.cnsns.2018.02.002}, \href
  {https://ui.adsabs.harvard.edu/abs/2018CNSNS..61..160P} {61, 160}

\bibitem[\protect\citeauthoryear{{Portinari}, {Chiosi}  \&
  {Bressan}}{{Portinari} et~al.}{1998}]{1998A&A...334..505P}
{Portinari} L.,  {Chiosi} C.,   {Bressan} A.,  1998, \aap, \href
  {https://ui.adsabs.harvard.edu/abs/1998A&A...334..505P} {334, 505}

\bibitem[\protect\citeauthoryear{{Pouteau} et~al.,}{{Pouteau}
  et~al.}{2022}]{2022arXiv220303276P}
{Pouteau} Y.,  et~al., 2022, arXiv e-prints, \href
  {https://ui.adsabs.harvard.edu/abs/2022arXiv220303276P} {p. arXiv:2203.03276}

\bibitem[\protect\citeauthoryear{{Prantzos} \& {Charbonnel}}{{Prantzos} \&
  {Charbonnel}}{2006}]{2006A&A...458..135P}
{Prantzos} N.,  {Charbonnel} C.,  2006, \mn@doi [\aap]
  {10.1051/0004-6361:20065374}, \href
  {https://ui.adsabs.harvard.edu/abs/2006A&A...458..135P} {458, 135}

\bibitem[\protect\citeauthoryear{{Reg{\H{o}}s}, {Tout}, {Wickramasinghe},
  {Hurley}  \& {Pols}}{{Reg{\H{o}}s} et~al.}{2003}]{2003NewA....8..283R}
{Reg{\H{o}}s} E.,  {Tout} C.~A.,  {Wickramasinghe} D.,  {Hurley} J.~R.,
  {Pols} O.~R.,  2003, \mn@doi [\na] {10.1016/S1384-1076(02)00221-X}, \href
  {https://ui.adsabs.harvard.edu/abs/2003NewA....8..283R} {8, 283}

\bibitem[\protect\citeauthoryear{{Renzini}, {Marino}  \& {Milone}}{{Renzini}
  et~al.}{2022}]{2022arXiv220303002R}
{Renzini} A.,  {Marino} A.~F.,   {Milone} A.~P.,  2022, arXiv e-prints, \href
  {https://ui.adsabs.harvard.edu/abs/2022arXiv220303002R} {p. arXiv:2203.03002}

\bibitem[\protect\citeauthoryear{{Schneider} et~al.,}{{Schneider}
  et~al.}{2018}]{2018Sci...359...69S}
{Schneider} F.~R.~N.,  et~al., 2018, \mn@doi [Science]
  {10.1126/science.aan0106}, \href
  {https://ui.adsabs.harvard.edu/abs/2018Sci...359...69S} {359, 69}

\bibitem[\protect\citeauthoryear{{Sharda} \& {Krumholz}}{{Sharda} \&
  {Krumholz}}{2022}]{2022MNRAS.509.1959S}
{Sharda} P.,  {Krumholz} M.~R.,  2022, \mn@doi [\mnras]
  {10.1093/mnras/stab2921}, \href
  {https://ui.adsabs.harvard.edu/abs/2022MNRAS.509.1959S} {509, 1959}

\bibitem[\protect\citeauthoryear{{Sharma}, {Corbelli}, {Giovanardi}, {Hunt}  \&
  {Palla}}{{Sharma} et~al.}{2011}]{2011A&A...534A..96S}
{Sharma} S.,  {Corbelli} E.,  {Giovanardi} C.,  {Hunt} L.~K.,   {Palla} F.,
  2011, \mn@doi [\aap] {10.1051/0004-6361/201117812}, \href
  {https://ui.adsabs.harvard.edu/abs/2011A&A...534A..96S} {534, A96}

\bibitem[\protect\citeauthoryear{Stein}{Stein}{1977}]{10.2307/2689506}
Stein S.~K.,  1977, Mathematics Magazine, 50, 160

\bibitem[\protect\citeauthoryear{{Sukhbold}, {Ertl}, {Woosley}, {Brown}  \&
  {Janka}}{{Sukhbold} et~al.}{2016}]{2016ApJ...821...38S}
{Sukhbold} T.,  {Ertl} T.,  {Woosley} S.~E.,  {Brown} J.~M.,   {Janka} H.~T.,
  2016, \mn@doi [\apj] {10.3847/0004-637X/821/1/38}, \href
  {https://ui.adsabs.harvard.edu/abs/2016ApJ...821...38S} {821, 38}

\bibitem[\protect\citeauthoryear{{Usher}, {Brodie}, {Forbes}, {Romanowsky},
  {Strader}, {Pfeffer}  \& {Bastian}}{{Usher}
  et~al.}{2019}]{2019MNRAS.490..491U}
{Usher} C.,  {Brodie} J.~P.,  {Forbes} D.~A.,  {Romanowsky} A.~J.,  {Strader}
  J.,  {Pfeffer} J.,   {Bastian} N.,  2019, \mn@doi [\mnras]
  {10.1093/mnras/stz2596}, \href
  {https://ui.adsabs.harvard.edu/abs/2019MNRAS.490..491U} {490, 491}

\bibitem[\protect\citeauthoryear{{Verliat}, {Hennebelle}, {Gonz{\'a}lez}, {Lee}
   \& {Geen}}{{Verliat} et~al.}{2022}]{2022arXiv220202237V}
{Verliat} A.,  {Hennebelle} P.,  {Gonz{\'a}lez} M.,  {Lee} Y.-N.,   {Geen} S.,
  2022, arXiv e-prints, \href
  {https://ui.adsabs.harvard.edu/abs/2022arXiv220202237V} {p. arXiv:2202.02237}

\bibitem[\protect\citeauthoryear{{Wang} \& {Jerabkova}}{{Wang} \&
  {Jerabkova}}{2021}]{2021A&A...655A..71W}
{Wang} L.,  {Jerabkova} T.,  2021, \mn@doi [\aap]
  {10.1051/0004-6361/202141838}, \href
  {https://ui.adsabs.harvard.edu/abs/2021A&A...655A..71W} {655, A71}

\bibitem[\protect\citeauthoryear{{Webb} \& {Bovy}}{{Webb} \&
  {Bovy}}{2021}]{2021arXiv210802217W}
{Webb} J.~J.,  {Bovy} J.,  2021, arXiv e-prints, \href
  {https://ui.adsabs.harvard.edu/abs/2021arXiv210802217W} {p. arXiv:2108.02217}

\bibitem[\protect\citeauthoryear{{Webb} \& {Leigh}}{{Webb} \&
  {Leigh}}{2015}]{2015MNRAS.453.3278W}
{Webb} J.~J.,  {Leigh} N. W.~C.,  2015, \mn@doi [\mnras]
  {10.1093/mnras/stv1780}, \href
  {https://ui.adsabs.harvard.edu/abs/2015MNRAS.453.3278W} {453, 3278}

\bibitem[\protect\citeauthoryear{{Willson}}{{Willson}}{2000}]{2000ARA&A..38..573W}
{Willson} L.~A.,  2000, \mn@doi [\araa] {10.1146/annurev.astro.38.1.573}, \href
  {https://ui.adsabs.harvard.edu/abs/2000ARA&A..38..573W} {38, 573}

\bibitem[\protect\citeauthoryear{{Wirth}, {Jerabkova}, {Yan}, {Kroupa}, {Haas}
  \& {{\v{S}}ubr}}{{Wirth} et~al.}{2021}]{2021MNRAS.506.4131W}
{Wirth} H.,  {Jerabkova} T.,  {Yan} Z.,  {Kroupa} P.,  {Haas} J.,
  {{\v{S}}ubr} L.,  2021, \mn@doi [\mnras] {10.1093/mnras/stab2011}, \href
  {https://ui.adsabs.harvard.edu/abs/2021MNRAS.506.4131W} {506, 4131}

\bibitem[\protect\citeauthoryear{{Yan}, {Jerabkova}  \& {Kroupa}}{{Yan}
  et~al.}{2017}]{2017A&A...607A.126Y}
{Yan} Z.,  {Jerabkova} T.,   {Kroupa} P.,  2017, \mn@doi [\aap]
  {10.1051/0004-6361/201730987}, \href
  {https://ui.adsabs.harvard.edu/abs/2017A&A...607A.126Y} {607, A126}

\bibitem[\protect\citeauthoryear{{Yan}, {Jerabkova}, {Kroupa}  \&
  {Vazdekis}}{{Yan} et~al.}{2019}]{2019A&A...629A..93Y}
{Yan} Z.,  {Jerabkova} T.,  {Kroupa} P.,   {Vazdekis} A.,  2019, \mn@doi [\aap]
  {10.1051/0004-6361/201936029}, \href
  {https://ui.adsabs.harvard.edu/abs/2019A&A...629A..93Y} {629, A93}

\bibitem[\protect\citeauthoryear{{Yan}, {Jerabkova}  \& {Kroupa}}{{Yan}
  et~al.}{2020}]{2020A&A...637A..68Y}
{Yan} Z.,  {Jerabkova} T.,   {Kroupa} P.,  2020, \mn@doi [\aap]
  {10.1051/0004-6361/202037567}, \href
  {https://ui.adsabs.harvard.edu/abs/2020A&A...637A..68Y} {637, A68}

\bibitem[\protect\citeauthoryear{{Yan}, {Je{\v{r}}{\'a}bkov{\'a}}  \&
  {Kroupa}}{{Yan} et~al.}{2021}]{2021A&A...655A..19Y}
{Yan} Z.,  {Je{\v{r}}{\'a}bkov{\'a}} T.,   {Kroupa} P.,  2021, \mn@doi [\aap]
  {10.1051/0004-6361/202140683}, \href
  {https://ui.adsabs.harvard.edu/abs/2021A&A...655A..19Y} {655, A19}

\bibitem[\protect\citeauthoryear{{Zhang}, {Romano}, {Ivison}, {Papadopoulos}
  \& {Matteucci}}{{Zhang} et~al.}{2018}]{2018Natur.558..260Z}
{Zhang} Z.-Y.,  {Romano} D.,  {Ivison} R.~J.,  {Papadopoulos} P.~P.,
  {Matteucci} F.,  2018, \mn@doi [\nat] {10.1038/s41586-018-0196-x}, \href
  {https://ui.adsabs.harvard.edu/abs/2018Natur.558..260Z} {558, 260}

\makeatother
\end{thebibliography}
